\newcommand{\eq}[1]{\begin{align}#1\end{align}}

\documentclass[twocolumn,aps,prb,superscriptaddress]{revtex4-1}
\usepackage{bm}
\usepackage{newtxmath}
\usepackage{array,booktabs}
\usepackage[dvipdfmx]{graphicx}
\usepackage{here}
\usepackage{color}

\begin{document}
%title
\title{Tunneling conductance in two-dimensional junctions between a normal metal and a ferromagnetic Rashba metal}

\author{Daisuke Oshima}
\affiliation{Department of Applied Physics, Nagoya University, Nagoya, 464-8603, Japan}

\author{Katsuhisa Taguchi}
\affiliation{Department of Applied Physics, Nagoya University, Nagoya, 464-8603, Japan}

\author{Yukio Tanaka}
\affiliation{Department of Applied Physics, Nagoya University, Nagoya, 464-8603, Japan}

%abstract
\begin{abstract}
We have studied charge transport in ferromagnetic Rashba metal (FRM),  
where  both Rashba type spin-orbit coupling (RSOC) and exchange coupling 
coexist. It  has nontrivial metallic states, $i.e.$, normal Rashba metal (NRM), anomalous Rashba metal (ARM), and Rashba ring metal (RRM), and they are manipulated by tuning the Fermi level with an applied gate voltage.
We theoretically studied tunneling conductance ($G$) in a normal metal / FRM junction by changing the Fermi level via an applied gate voltage ($V_g$) on the FRM. We found a wide variation in the $V_{g}$ dependence of $G$, which depends on the metallic states.
In NRM, the $V_g$ dependence of $G$ is the same as that in a conventional two-dimensional system. 
However, in ARM, the $V_g$ dependence of $G$ is similar to that in a conventional one (two)-dimensional system for a large (small) RSOC.
Furthermore, in RRM, which is generated by a large RSOC,  
the $V_g$ dependence of the $G$ is similar to that in the one-dimensional system. 
In addition, these anomalous properties stem from the density of states in ARM and RRM caused by the large RSOC and exchange coupling rather than the spin-momentum locking of RSOC.
\end{abstract}

\maketitle

%------------------------------------------
%Introduction      
%------------------------------------------
\section{Introduction}
\begin{figure}[htbp]
\centering
\includegraphics[width = 85mm]{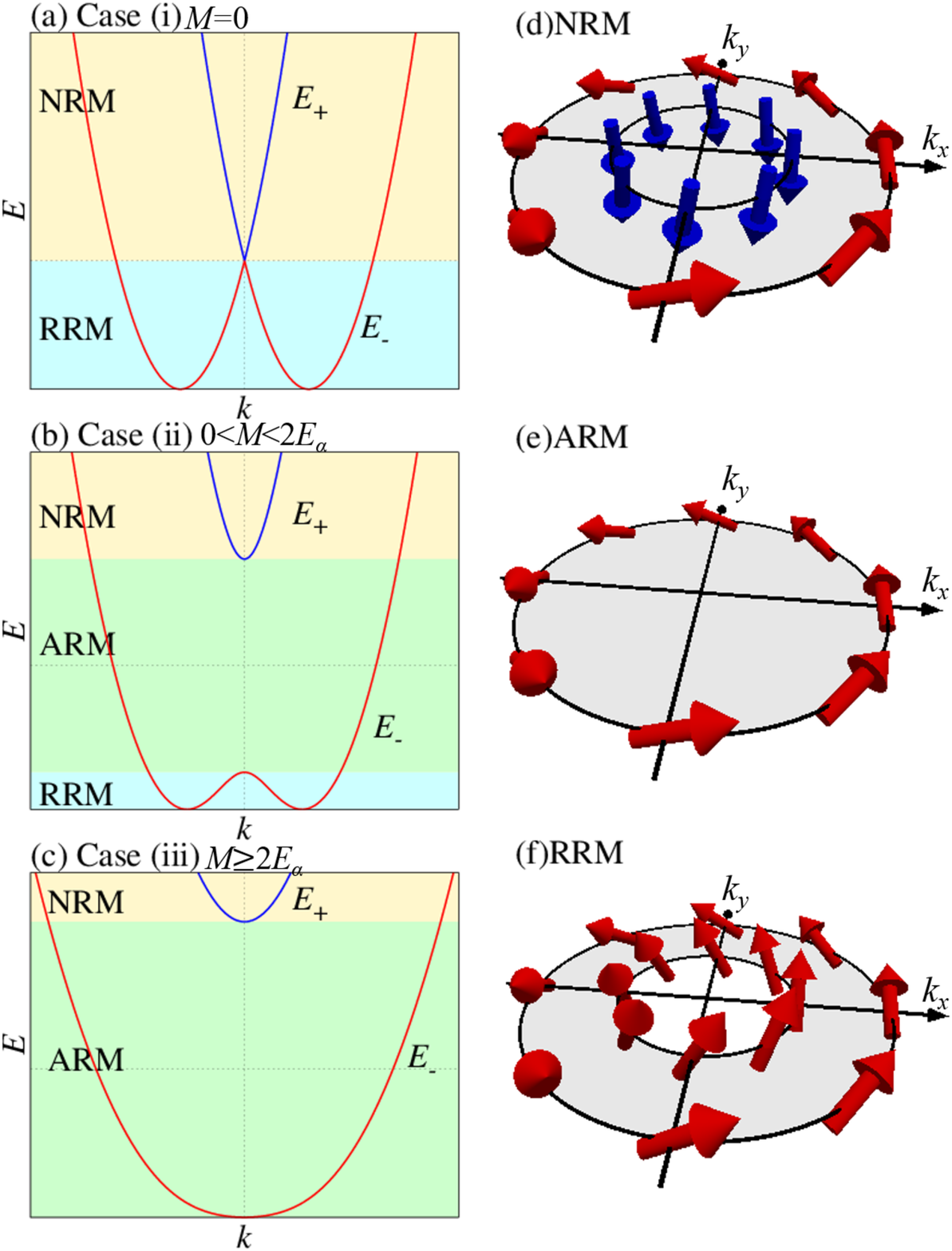}
\caption{(Color Online) The spin-split energy dispersion of the FRM in (a)$M=0$, (b)$0<M<2 E_\alpha$, and (c)$M \ge 2E_{\alpha}$, where $E_{\alpha}=m_{\rm F}\alpha^2/(2\hbar^2)$ and $M$ denote the Rashba energy and exchange coupling, respectively. 
The yellow, green, and blue regions correspond to NRM, ARM, and RRM, respectively. 
RRM appears when $2E_\alpha >M$ is satisfied.
The Fermi surface in these metallic states for $M\neq 0$ is illustrated in Figs. (d)-(f), where the arrows denote the spin polarization in the spin space at each $\bm{k}$ and gray regions show the regions of the occupied states in the momentum space.
}
\label{structure}
\end{figure}

Rashba type spin-orbit coupling (RSOC), which has promising potential for controlling charge transport, has properties of spin-momentum locking and breaking the spin degeneracy of energy bands. 
The former properties become prominent on the surface of topological insulators (TI) \cite{Yokoyama10,Mondal10,Taguchi14}. 
The latter properties directly reflect on the charge transport in tunneling junctions \cite{Molenkamp01,Matsuyama02,Jiang03,Ramaglia03,Yokoyama06,Srisongmuang08,Matos-Abiague09,Zhang13,Jantayod13,Jantayod15}.
For example, a metallic junction with RSOC provides an intriguing tunneling conductance, which depends on the applied gate voltage ($V_g$), namely, the position of the Fermi level \cite{Srisongmuang08,Jantayod13,Jantayod15}.
The $V_{g}$ dependence of conductance $G(V_g)$ is based on whether the Fermi level crosses two spin-split bands $E_{+}$ and $E_{-}$, or only $E_{-}$ [see in Fig. \ref{structure}(a)]\cite{Jantayod13,Jantayod15}.
Current topic of spintronics is to study charge transport in the presence of spin-orbit coupling with magnetization or an applied magnetic field \cite{Grundler01,Larsen02,Streda03,Krupin05,Sanchez08,Fallahi12,Pang12,Tang12,Tang16,Fukumoto15,Han15,Wojcik15}.

%---------------------------
% Properties in FRM 
%---------------------------
%Figure
%Figure end
%
The simultaneous existence of RSOC and exchange coupling of magnetization ($M$) generates three types of metallic states. 
We call this metallic system a ferromagnetic Rashba metal (FRM). 
These three states have different configurations at the Fermi surface.
(i) The first case appears when the Fermi level crosses the two bands $E_\pm$.
In this case, there are inner and outer spin-dependent Fermi surfaces, as shown in Fig. \ref{structure}(d). 
We call this normal Rashba metal (NRM).
(ii) The second state is realized when the Fermi level crosses only the $E_-$ band. 
The number of spin-split Fermi surfaces becomes one, as shown in Figs. \ref{structure}(b)-(c) and Fig. \ref{structure}(e).
This state is called anomalous Rashba metal (ARM) \cite{Fukumoto15}. 
(iii) The third state occurs when the Fermi level is located below $E_-(\bm{k}=0)$. 
We call this states Rashba ring metal (RRM).
Remarkably, the shape of the region of the occupied states in the momentum space in the RRM is different from that in the NRM as well as the ARM; the corresponding regions in RRM and NRM are ring- and disc-shaped, respectively.
Therefore, the energy dependence of the DOS is dramatically different for each state, and it is expected that changing the states could affect physical phenomena directly. 

It should be noted that in the presence of magnetization, the RRM appears in the case where the energy scale of the RSOC is larger than that of $M$ [see Figs. \ref{structure}(b) and \ref{structure}(c)]. 
%
%---------------------------
% Such a situation can be realized in  
%---------------------------
Although the presence of the RRM requires the system to host a large RSOC, recent experiments have been reported on two-dimensional (2D) systems with large RSOC and magnetization, e.g., the heterostructure of Pt/Co/Al-oxides \cite{Miron10}.
Therefore, the RRM and ARM in a FRM can be realized by tuning the Fermi level with an applied gate voltage in thin layered heterostructures.

%------------------------------------------
% This work         
%------------------------------------------
In this paper, we describe the gate voltage dependence of the tunneling conductance $G(V_g)$ in a 2D normal metal (NM)/FRM tunneling junction, when the magnetization of the FRM is along the out-of-plane direction. 
We found a wide variation in the $V_{g}$ dependence of the conductance in NRM, ARM, and RRM. 
In particular, although we studied 2D systems, 
the obtained results in ARM and RRM are similar to those in conventional one-dimensional (1D) systems.
It is noted that the 1D-like features emerge in the 2D junctions by tuning the Fermi level.
Additionaly, we clarified that the anomalous properties stem from the DOS in ARM and RRM rather than the spin-momentum locking of RSOC.
These results could be useful when we use materials in ARM and RRM.

%------------------------------------------
% Organization       
%------------------------------------------
The organization of this paper is as follows. 
In Sect. \ref{sec:II}, we present FRM, a model of the NM/FRM junctions and a method to calculate $G(V_g)$.
In Sect. \ref{sec:III}, we show the $V_g$ dependence of $G$ in various cases. 
In Sect. \ref{sec:IV}, to explain the origin of the anomalous $V_g$ dependence, 
we discuss the systematic change of the DOS in NRM, ARM, and RRM. 
In Sect. \ref{sec:V}, we summarize the results and discuss a way to experimentally detect the anomalous tunneling conductance.

\section{Model}\label{sec:II}
\subsection{Ferromagnetic Rashba metal (FRM)}
We introduce a system of FRM.
Its effective  Hamiltonian is described as \cite{Rashba60,Cayao15,Streda03,Fukumoto15} 
\eq{
{H}_{{\rm FRM}}&= \frac{\hbar^2k^2}{2m_{\rm F}}+\alpha {(\bm{k} \times\bm{\sigma})}_z-M \sigma_z,
\label{HamiltonianNM/FRM}
}
with $\bm{k}=(k_x, k_y)$ and $k^2= k_x^2+k_y^2$.
Here, $m_{\rm F}$ is the effective mass of the FRM.
The first term expresses the kinetic energy. 
The second term denotes the RSOC; 
$\alpha$ is the strength of the RSOC. 
The third term $M \sigma_z$ denotes the exchange coupling.
${\bm \sigma}$ is the Pauli matrices in spin space.

From this Hamiltonian, the energy dispersion becomes 
\eq{
E_\pm(k)=\frac{\hbar^2 k^2 }{2m_{\rm F}} \pm \sqrt{ \alpha ^ 2 k^2 + M^2}  
\label{blanches}.
}
The dispersion can be classified into three cases: (i) $M=0$, (ii) $0<M<2E_\alpha$, and (iii) $M\geq 2E_\alpha$ as shown in Figs. \ref{structure}(a)-(c).
Here, $E_\alpha =m_{\rm F}\alpha^2/(2\hbar^2)$ is a Rashba energy\cite{Ast07,Sablikov07,Srisongmuang08,Ast08,Mathias10,Ishizaka11,Jantayod13,Cayao15}.
In case (i), RSOC lifts the spin degeneracy except for $k=0$\cite{Reeg17}, and the energy dispersion splits into two branches.
The $E_-$ branch has an annular minimum \cite{Srisongmuang08} for $k=\sqrt{2m_{\rm F}E_\alpha}/\hbar$.
In case (ii), the two branches perfectly split because of the presence of $M$.
The $E_-$ branch has an annular minimum for $k=\sqrt{2m_{\rm F}(E_\alpha-E_c)}/\hbar$, which is affected by $E_c=\hbar^2 M^2/(2m_{\rm F}\alpha^2)$.
In case (iii), the $E_-$ branch has a minimum at $\bm{k}=0$.

The spin structure at the Fermi surface depends on the position of the Fermi level, because of the RSOC and $M$.
As a result, generated spin structures are specific for  NRM, ARM, and RRM [see Figs. \ref{structure}(d)-(f)].
In NRM, there are two Fermi surfaces with almost opposite spin directions\cite{Streda03}. 
In ARM, the inner Fermi surface disappears.
In RRM, the inner and outer Fermi surfaces take the almost same spin structure.

\subsection{NM/FRM junction}
\begin{figure}[htbp]
\begin{center}
\includegraphics[width=85mm]{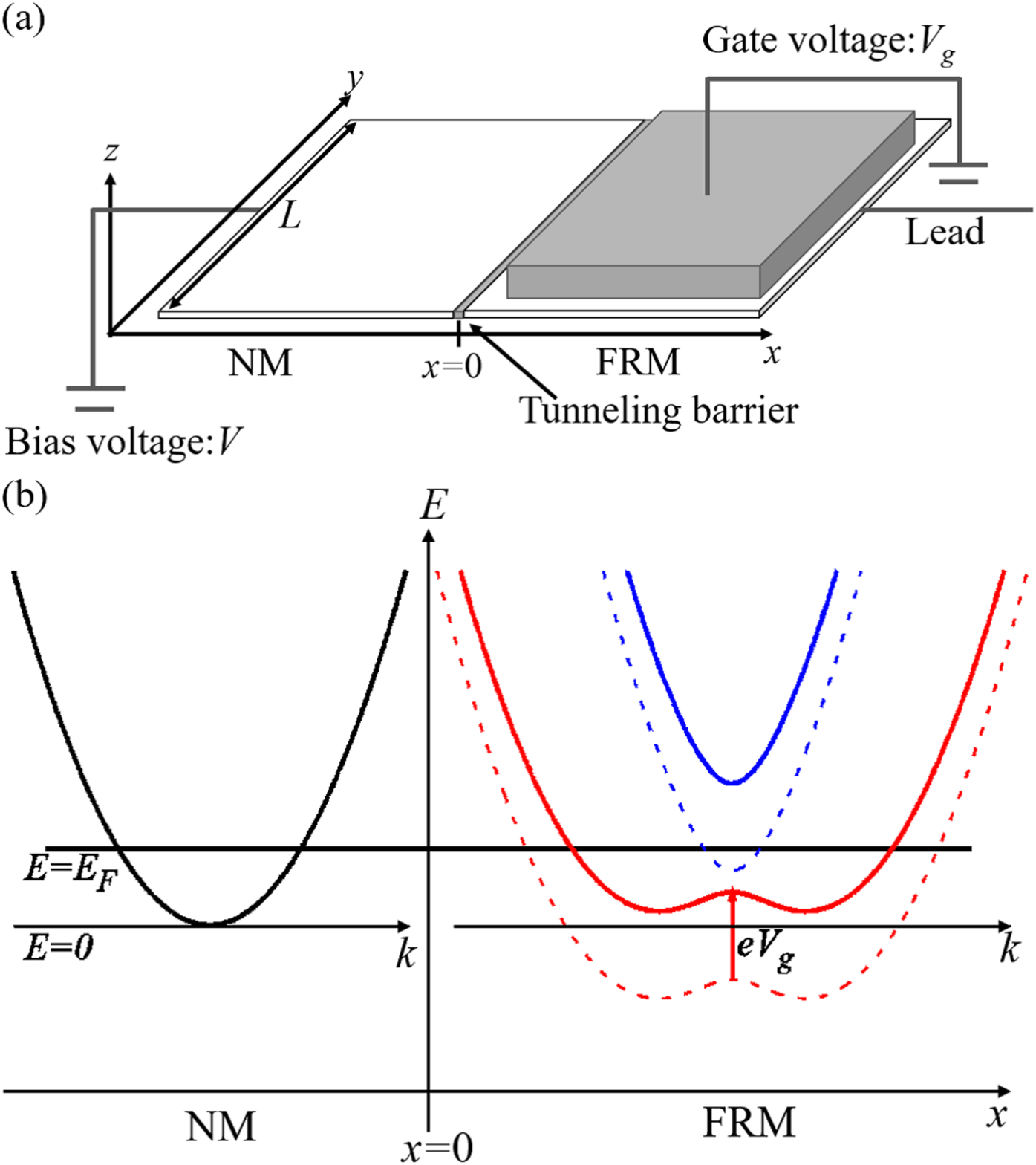}
\end{center}
\caption{(Color Online) (a) Schematic of the NM/FRM junction. The NM is applied by a bias voltage ($V$), and the FRM is applied by a gate voltage ($V_g$). $L$ is the width of the junction along $y$ direction.
(b) The energy dispersion of the NM (left panel) and FRM (right panel) in the junction. 
We define the Fermi level $E_{F}$ as measured from the bottom of the dispersion of the NM.
In the presence of the gate voltage $V_g$ in the right side, the Fermi level of the FRM is shifted by $V_g$.
In other words, in the existence of $V_g$, the energy band dispersion (dashed lines) changes to the solid lines.
}
\label{junctionmodel}
\end{figure}

We consider the tunneling conductance $G$ in NM/FRM junctions.
In the FRM region (right side of the junction), both RSOC and exchange coupling exist; 
We assume that a nonzero RSOC is generated by inversion symmetry breaking along the out-of-plane direction because of an internal \cite{Miron10} or external electric field of the applied gate voltage $V_g$ \cite{Datta90} [see Fig. \ref{junctionmodel}(a)].
Exchange coupling $M$ is caused by the magnetization, which is along the out-of-plane direction.
The Fermi level in FRM can be tuned by using the applied gate voltage $V_g$.
 The gate voltage plays a role in shifting the spin-split energy band of FRM $E_\pm$, as illustrated in Fig. \ref{junctionmodel} (b).
This gate voltage is assumed to affect only the FRM.
On the left side of the junction, RSOC, $M$, and $V_g$ are absent.
A bias voltage ($V$) is applied on the NM, and its magnitude is much smaller than that of Fermi energy $E_F$.
At the interface between NM and FRM, a delta-function type of tunneling barrier is assumed\cite{Srisongmuang08,Jantayod13,Fukumoto15}. 
It is also supposed that the interface is ideally a flat interface, and the $y$ component of momentum of the wave function is conserved. 

%Figure
%Figure end

The Hamiltonian model in the junction can be described as 
\eq{ \label{eq:total}
H=\frac{\hbar^2k^2}{2m_{\rm N}}\theta(-x) + U\delta(x) + ({H}_{{\rm FRM}}+eV_g)\theta(x),
}
where $\theta$ and $\delta$ are the Heaviside step function and delta function, respectively. 
The first term in Eq. (\ref{eq:total}) denotes the kinetic energy 
of the left side of the junction ($x<0$) given by $\hbar^2k^2/(2m_{\rm N})$, and $m_{\rm N}$ is the effective mass of the NM.
The second term $U\delta(x)$ indicates the tunneling barrier.
$U$ is the strength of the tunneling barrier.
The third term $H_{\rm FRM}$ expresses the effective Hamiltonian of FRM with the gate voltage in the right side of the junction ($x>0$). 
$eV_g\geq 0$ is a potential caused by the gate voltage, where $e$ is the elementary charge of an electron.
The relation between $V_g$ and the energy band of FRM are described in Fig. \ref{junctionmodel} (b). 
We assume a periodic boundary conditions along $y$ direction, 
and we set $\alpha>0$, and $M\ge0$.
Here, $L$ is the width of the junction along the $y$ direction.
We assume that $L$ is sufficiently large, and $k_y$ is a good quantum number.

\subsection{Conductance}
%Figure
\begin{figure*}[thb]
\begin{center}
\includegraphics[width=180mm]{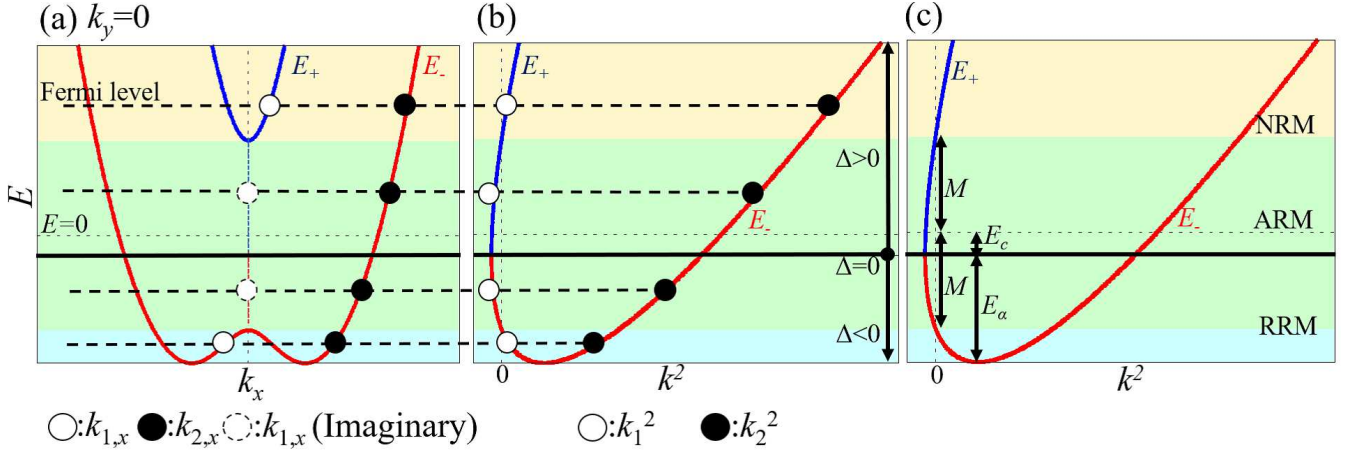}
\end{center}
\caption{(Color Online) 
Relation between the momentum of the transmitted wave function ${\bm k}_1$ (open circle) and ${\bm k}_2$ (closed circle) of each Fermi level (dashed line). 
(a) $k_{1,x}$ and $k_{2,x}$ at $k_y=0$ in the band structure. 
We find that in ARM (colored green region), $k_{1,x}$ takes purely imaginary number (dashed closed circle), since $k_{1,x}^2=k_1^2-k_{y}^2<0$.
(b) Energy dispersion of the FRM plotted as a function of $k^2$.
It is observed that $k_1^2 (<k^2_2)$ takes a negative value in the ARM of each Fermi level\cite{Sablikov07}. Moreover, we find that $k_1^2$ is given by $E_+$ for $\Delta>0$ and $E_-$ for $\Delta<0$; Therefore, the wave function corresponding to ${\bm k}_1$ is changed at $\Delta=0$ (bold line).
(c) Exchange coupling $M$, Rashba energy $E_\alpha =m_{\rm F}\alpha^2/(2\hbar^2)$, and 
$E_c=\hbar^2 M^2/(2m_{\rm F}\alpha^2)$.}
\label{k2dispersion}
\end{figure*}
%Figure end

To obtain the conductance, we consider the scattering process at the interface.
From Eq. (\ref{HamiltonianNM/FRM}), we obtain the wave function in the left side of the junction $\psi^{\uparrow (\downarrow)}(x<0,y)$.
Hereafter, the superscript $\uparrow(\downarrow)$ denotes the up (down) spin injection. 
$\psi^{\uparrow (\downarrow)}(x<0,y)$ is decomposed into the injected wave function $\psi^{\uparrow(\downarrow)}_{\textrm{in}}$,
and reflects one $\psi^{\uparrow(\downarrow)}_{\textrm{ref}}$ as 
\eq{ \label{eq:left}
&\psi^{\uparrow (\downarrow)}(x<0,y)=\psi_{\textrm{in}}^{\uparrow (\downarrow)}+\psi_{\textrm{ref}}^{\uparrow (\downarrow)},\\
&\psi_{\textrm{in}}^{\uparrow}=e^{i(k_x x+k_y y)}\begin{pmatrix}1 \\ 0\end{pmatrix},\quad\psi_{\textrm{in}}^{\downarrow}=e^{i(k_x x+k_y y)}\begin{pmatrix}0 \\ 1\end{pmatrix}\label{eq:in},\\
&\psi_{\textrm{ref}}^{\uparrow (\downarrow)}=e^{i(-k_x x+k_y y)}\begin{pmatrix}r^{\uparrow(\downarrow)}_\uparrow \\ r^{\uparrow(\downarrow)}_\downarrow\end{pmatrix},
\label{eq:ref}
}
with $k_x=k\cos\phi$ and $k_y=k\sin\phi$.
Here, $k=\sqrt{2m_{\rm N}E}/\hbar$ and $\phi=\tan^{-1}(k_y/k_x)$ are the momentum of the electron in $x<0$ and 
the angle between $\bm{k}$ and the $x$ axis, respectively.
$r^{\uparrow (\downarrow)}_\uparrow$ [$r^{\uparrow (\downarrow)}_\downarrow$] is the reflection coefficient of up [down] spin electron with up (down) spin injection, and it includes the spin-flip process. 

The transmitted wave function $\psi^{\uparrow (\downarrow)}(x>0,y)\equiv\psi_{\textrm{tra}}^{\uparrow (\downarrow)}$ is characterized by $\Delta\equiv E-eV_g+E_c $ as 
\eq{
&\psi_{\rm tra}^{\uparrow (\downarrow)}=t_1^{\uparrow (\downarrow)} \left[\theta(\Delta)\chi_+(\bm{k}_1)+\theta(-\Delta)\chi_-(\bm{k}_1)\right]+t_2^{\uparrow (\downarrow)}\chi_-(\bm{k}_2),
\label{transmissionwave}
}
with
\eq{
\chi_+(\bm{k})=e^{i\bm{k}\cdot\bm{r}}\begin{pmatrix} g_-(\bm{k}) \\ 1 \end{pmatrix}, \quad\chi_-(\bm{k})=e^{i\bm{k}\cdot\bm{r}}\begin{pmatrix} 1 \\ g_+(\bm{k}) \end{pmatrix}.
}
Here, $t_1^{\uparrow (\downarrow)}$ $[t_2^{\uparrow (\downarrow)}]$ denotes 
the transmission coefficient with up (down) spin injection. 
$\bm{k}_1 = (k_{1,x},k_y )$ and $\bm{k}_2 = (k_{2,x},k_y )$ are the momentum in FRM, which are defined for $k_1^2\leq k_2^2$ with $k_{1(2)}^2=k_{1(2),x}^2+k_y^2$. $\bm{k}_1$ and $\bm{k}_2$ correspond to the inner and outer Fermi surface, respectively. 
$\chi_\pm$ is the eigenfunction for the eigenvalue $E_\pm$.
We set $g_\pm(\bm{k}_{1(2)})=-\alpha i\left(k_{1(2),x}\pm ik_y\right)/\left(M+\sqrt{ \alpha^2 k_{1(2)}^2+M^2}\right)$. 
From the energy dispersion $E_\pm(k)$, $k_{1(2)}^2$ is given by
\eq{
k_{1(2)}^2&=\frac{2m_{\rm F}}{\hbar^2}\left[(E-eV_g)+2E_\alpha\right.\nonumber\\
&\qquad\qquad\left.-(+)\sqrt{4E_\alpha (E-eV_g)+4E_\alpha^2+M^2}\right].
\label{k1(2)}
}
Here, the energy $E$ in the FRM is shifted by the gate voltage $V_g$.

It is noted that an evanescent wave occurs because $k_{1,x}^2<0$ is negative in ARM. 
$E_+$ and $E_-$ cross $\Delta =0$, and $k_1^2$ is given by $E_+$ for $\Delta>0$ and $E_-$ for $\Delta<0$, as shown in Figs. \ref{k2dispersion}(b) and \ref{k2dispersion}(c).
For this fact, in ARM, the evanescent wave corresponding to ${\bm k}_1$ is described as $\chi_+({\bm k}_1)$ for $\Delta >0$ and $\chi_-({\bm k}_1)$ for $\Delta<0$ as shown in Eq. (\ref{transmissionwave}).
A short summary of ${\bm k}_1$ and ${\bm k}_2$ for each metallic state is presented in Table \ref{t1}.

To obtain $k_{1,x}$ and $k_{2,x}$, we consider the velocity operator $v_x$. 
Because an electron is injected along the $+x$ direction, the velocity $v_x$ should take a positive value, where 
the velocity $v_x=\partial H/(\hbar\partial k_x)$ is given by Eq. (\ref{HamiltonianNM/FRM}) as \cite{Streda03,Srisongmuang08} 
\eq{
v_x &= \frac{\hbar k_x }{m(x)} + \frac{\alpha}{\hbar} \theta(x) \sigma_y,\\
m(x)&=m_{\rm N}\theta(-x)+m_{\rm F}\theta(x).\nonumber
\label{velocity}
}
When $k_{1,x}$ ($k_{2,x}$) becomes an imaginary number, its sign is determined so that $\chi_{\pm} \to 0$ in the limit of $x \to \infty$.

%Table
\begin{table}[t]
\centering
\caption{Origin of the momentum of transmitted wave
${\bm k}_1=(k_{1,x},k_y)$ and ${\bm k}_2=(k_{2,x},k_y)$ for NRM, ARM, and RRM.
$\Delta$ equals $E-eV_g+E_c$.}
\begin{tabular}{ccc}
\hline
State& $E_+$ & $E_-$ \\
\hline
NRM & ${\bm k}_1$ & ${\bm k}_2$\\
ARM $[\Delta>0]$ &${\bm k}_1$ &${\bm k}_2$\\
ARM $[\Delta<0]$ &--&${\bm k}_1$, ${\bm k}_2$\\
RRM &--&${\bm k}_1$, ${\bm k}_2$\\
\hline
\end{tabular}
\label{t1}
\end{table}
%Table end

To solve the wave function, we consider the boundary condition at the interface \cite{Molenkamp01,Srisongmuang08,Zulicke01,Jantayod13,Fukumoto15,Reeg17}:
\begin{align}
\begin{aligned}
&\psi^{\uparrow (\downarrow)}(+0,y)-\psi^{\uparrow (\downarrow)}(-0,y)=0,\\
&v_x[\psi^{\uparrow (\downarrow)}(+0,y)-\psi^{\uparrow (\downarrow)}(-0,y)]=\frac{2U}{i\hbar}\psi^{\uparrow (\downarrow)}(0,y).
\end{aligned}
\end{align}
Then, we obtain the probability current density $j_x(k_y)={\rm Re}[\psi^{\uparrow (\downarrow)\dagger} \sl{v_x}\psi^{\uparrow (\downarrow)}]$ \cite{Zulicke01,Srisongmuang08,Jantayod13}, the reflection probability $R^{\uparrow (\downarrow)}$, and the transmission probability $T^{\uparrow (\downarrow)}$ given by
\begin{align}
\begin{aligned}
R^{\uparrow (\downarrow)}(E,\phi)=&\left|\frac{j^{\uparrow (\downarrow)}_{x, \textrm{ref}}}{j^{\uparrow (\downarrow)}_{x,\textrm{in}}}\right|
	={\rm Re}\left|\frac{\psi^{\uparrow (\downarrow)\dagger}_{\textrm{ref}}v_x\psi^{\uparrow (\downarrow)}_{\textrm{ref}}}{\psi^{\uparrow (\downarrow)\dagger}_{\textrm{in}} v_x\psi^{\uparrow (\downarrow)}_{\textrm{in}}}\right|,\\
T^{\uparrow (\downarrow)}(E,\phi)=&\left|\frac{j^{\uparrow (\downarrow)}_{x,\textrm{tra}}}{j^{\uparrow (\downarrow)}_{x,\textrm{in}}}\right|
	={\rm Re}\left|\frac{\psi^{\uparrow (\downarrow)\dagger}_{\textrm{tra}} v_x\psi^{\uparrow (\downarrow)}_{\textrm{tra}}}{\psi^{\uparrow (\downarrow)\dagger}_{\rm in} v_x\psi^{\uparrow (\downarrow)}_{\textrm{in}}}\right|.
\end{aligned}
\end{align}
Here, $j^{\uparrow (\downarrow)}_{x,\textrm{in}}$, $j^{\uparrow (\downarrow)}_{x,\textrm{ref}}$, and $j^{\uparrow (\downarrow)}_{x,\textrm{tra}}$ are the $x$ component of the injected, reflected, and transmitted 
probability current density, respectively. 
From $R^{\uparrow (\downarrow)}(E,\phi)+T^{\uparrow (\downarrow)}(E,\phi)=1$, we can obtain $T^{\uparrow (\downarrow)}$ as 
\eq{
T^{\uparrow (\downarrow)}(E,\phi)=1-\left({|r^{\uparrow (\downarrow)}_\uparrow|}^2+{|r^{\uparrow (\downarrow)}_\downarrow|}^2\right).
\label{T1}
}
Finally, since the bias voltage $V$ is very weak, at the low-temperature limit, the electric current $I$ 
from the left lead to the right lead is given as \cite{Sablikov07,Srisongmuang08,Jantayod13}:
\begin{align}
I&=\frac{eL}{4\pi^2\hbar}\int^\infty_{-\infty} dE\int_{-\pi/2}^{\pi/2} d\phi \cos\phi \cdot k \left[T^\uparrow(E,\phi)+T^\downarrow(E,\phi)\right]\nonumber\\
 &\qquad\qquad\qquad\qquad\times\left[f(E-E_F-eV)-f(E-E_F)\right]\nonumber\\
 &\approx \frac{e^2VL}{4\pi^2\hbar}\int_{-\pi/2}^{\pi/2} d\phi \cos\phi \cdot k_F[T^\uparrow(E_F,\phi)+T^\downarrow(E_F,\phi)].
\end{align}
where $k_F= \sqrt{2m_{\rm N}E_F}/\hbar$, and $f(E-E_F)$ are the Fermi momentum in the NM, and the Fermi distribution function, respectively.
Here, we use $f(E-E_F-eV)-f(E-E_F)\approx eV\delta (E-E_F)$.
At the zero bias limit, the tunneling conductance per unit width, $G=dI/(LdV)$, is given by
\begin{align}
\label{eq:conductivity-NM-FRM}
G = \frac{e^2}{4\pi^2\hbar}\int_{-\pi/2}^{\pi/2} d\phi \cos\phi \ \cdot k_F\left[T^\uparrow(E_F,\phi)+T^\downarrow(E_F,\phi)\right].
\end{align}

%------------------------------------------
%Results
%------------------------------------------
\section{Result}
\label{sec:III}
%Figure
\begin{figure}[h]
\begin{center}
\includegraphics[width=85mm]{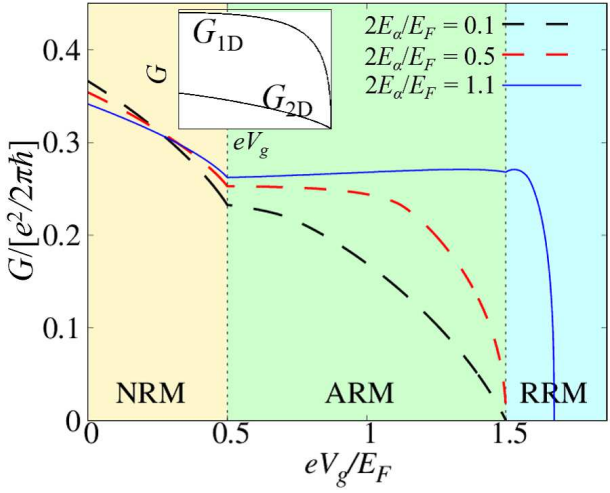}
\end{center}
\caption{(Color Online) $V_g$ dependence of $G$ for several $E_\alpha$ values at $M/E_F=0.5$, $Uk_F/E_F=1.0$, and $m_{\rm F}/m_{\rm N}=1$ with $k_F=\sqrt{2m_{\rm N}E_F}/\hbar$.
Here, $E_\alpha$, $M$, and $U$ are the Rashba energy, exchange coupling, and strength of the potential barrier, respectively.
In this case, $2E_\alpha /E_F=0.1, 0.5$ and $2E_\alpha /E_F=1.1$ show the band structure of FRM in cases (iii) and (ii), respectively [see Figs. \ref{structure}(b)-(c)].
(inset) $G_{\rm 1D}$ and $G_{\rm2D}$ show a typical $V_g$ dependence of tunneling conductance in 1D and 2D NM/NM junctions, respectively.
$G_{\rm 1D}$ is almost independent of $V_g$ except around the band bottom.  $G_{\rm 1D}$ decreases rapidly with increasing $V_g$ and its slope becomes divergent when the Fermi level is located near the band bottom.
In addition, $G_{\rm 2D}$ decreases monotonically with increasing $V_g$.
These features are independent of $m_{\rm F}/m_{\rm N}$.
}
\label{NMFRMG_a}
\end{figure}
%Figure end

First, we assume that the effective mass of each side is equal,  $m_{\rm F}/m_{\rm N}=1$.
Figure \ref{NMFRMG_a} shows the gate voltage dependence of the tunneling conductance $G(V_g)$ in the NM/FRM junction for several $E_\alpha$ values at $Uk_F/E_F=1.0$ with $k_F=\sqrt{2m_{\rm N}E_F}/\hbar$. 
To clarify properties of $G(V_g)$, we consider a typical tunneling conductance $G_{\rm 1D}$  ($G_{\rm 2D}$) in the absence of RSOC and magnetization in a 1D (2D) NM/NM tunneling junction. 
In these junctions, the gate voltages $V_g$ are attached to the right sides of the junctions.
Then, it is known that $G_{\rm 1D}$ is almost independent of $V_g$ except when the Fermi level is located far from the band bottom on the right side.
For the large magnitude of $V_g$,
$G_{\rm 1D}$ decreases rapidly with increasing $V_g$ and its slope becomes divergent when the Fermi level is near the band bottom on the right side.
$G_{\rm 2D}$ decreases monotonically (see inset in Fig. \ref{NMFRMG_a}).
We also found that these features of $G_{\rm 1D}$ and $G_{\rm 2D}$ are independent of $m_{\rm F}/m_{\rm N}$.

We find that, in NRM ($0< eV_g/E_F < 0.5$), $G$ decreases monotonically with increasing $V_g$, 
and its $V_g$ dependence is almost independent of $E_\alpha$. 
Such a monotonic dependence is the same as that of $G_{\rm 2D}$.
In ARM ($0.5< eV_g/E_F<1.5$), the qualitative feature 
of the $G(V_g)$ depends on whether $2E_\alpha > M$ is satisfied.
When $2E_\alpha$ is smaller than $M$ ($2E_\alpha /E_F=0.1$ in Fig. \ref{NMFRMG_a}), 
$G$ decreases monotonically with increasing $V_g$.
This $V_g$ dependence is similar to that of $G_{\rm 2D}$. 
However, when $2E_\alpha >M$ is satisfied ($2E_\alpha /E_F =1.1$ in Fig. \ref{NMFRMG_a}), 
$G$ is almost independent of $V_g$.
This behavior is similar to that of $G_{\rm 1D}$.
In RRM ($1.5< eV_g/E_F$), near the band bottom of FRM, the $V_g$ dependence of the $G$ is similar to 
that of $G_{\rm 1D}$.
Even in the 2D junction, the conductance in ARM is 1D-like (2D-like) for strong (weak) RSOC. 
In addition, in RRM, the $V_g$ dependence is similar to that of $G_{\rm 1D}$.
It is noted that we consider only the line shape of the $V_g$ dependence of $G$. The obtained results do not imply that the actual motion of an electron is 1D-like or 2D-like.

%Figure
\begin{figure}[h]
\begin{center}
\includegraphics[width=85mm]{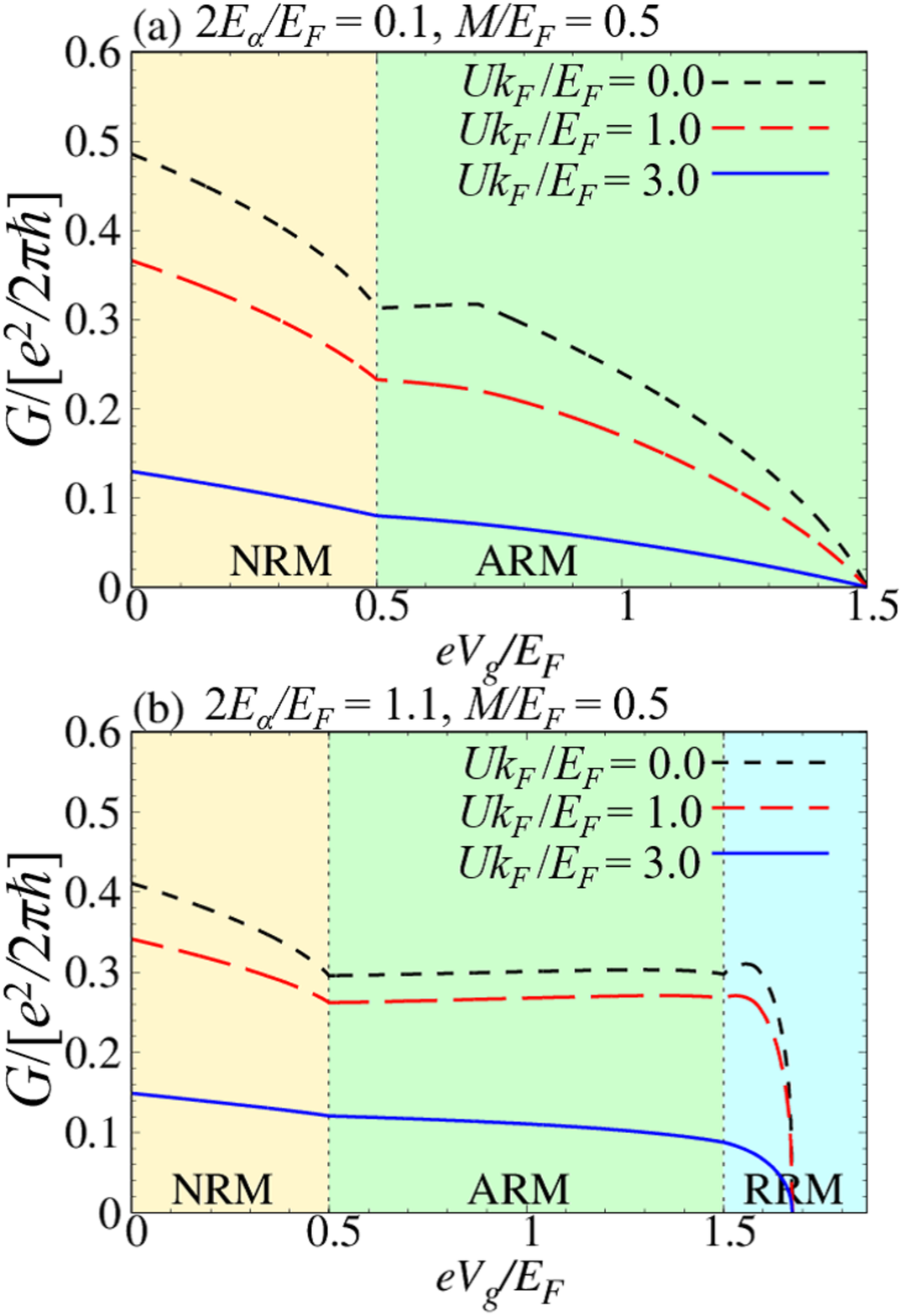}
\end{center}
\caption{(Color Online) $V_g$ dependence of $G$ for several $U$ values with $M/E_F=0.5$ and $m_{\rm F}/m_{\rm N}=1$ in (a) weak RSOC ($2E_\alpha /E_F=0.1$) and (b) strong RSOC ($2E_\alpha /E_F=1.1$).
(a) and (b) correspond to the band structure of Figs. \ref{structure}(c) and \ref{structure}(b), respectively.}
\label{NMFRMG_V}
\end{figure}
%Figure end

Figure \ref{NMFRMG_V}(a) and \ref{NMFRMG_V}(b) show the $V_g$ dependence of $G$ for any strength of the potential barrier: $Uk_F/E_F=0, 1, 3$, in the presence of weak RSOC ($2E_\alpha <M: 2E_\alpha /E_F=0.1, M/E_F=0.5$) and in strong RSOC ($2E_\alpha >M: 2E_\alpha /E_F=1.1, M/E_F=0.5$).
Both figures show that the $V_g$ dependence is qualitatively the same for any strength of the potential barrier.

%Figure
\begin{figure}[h]
\begin{center}
\includegraphics[width=85mm]{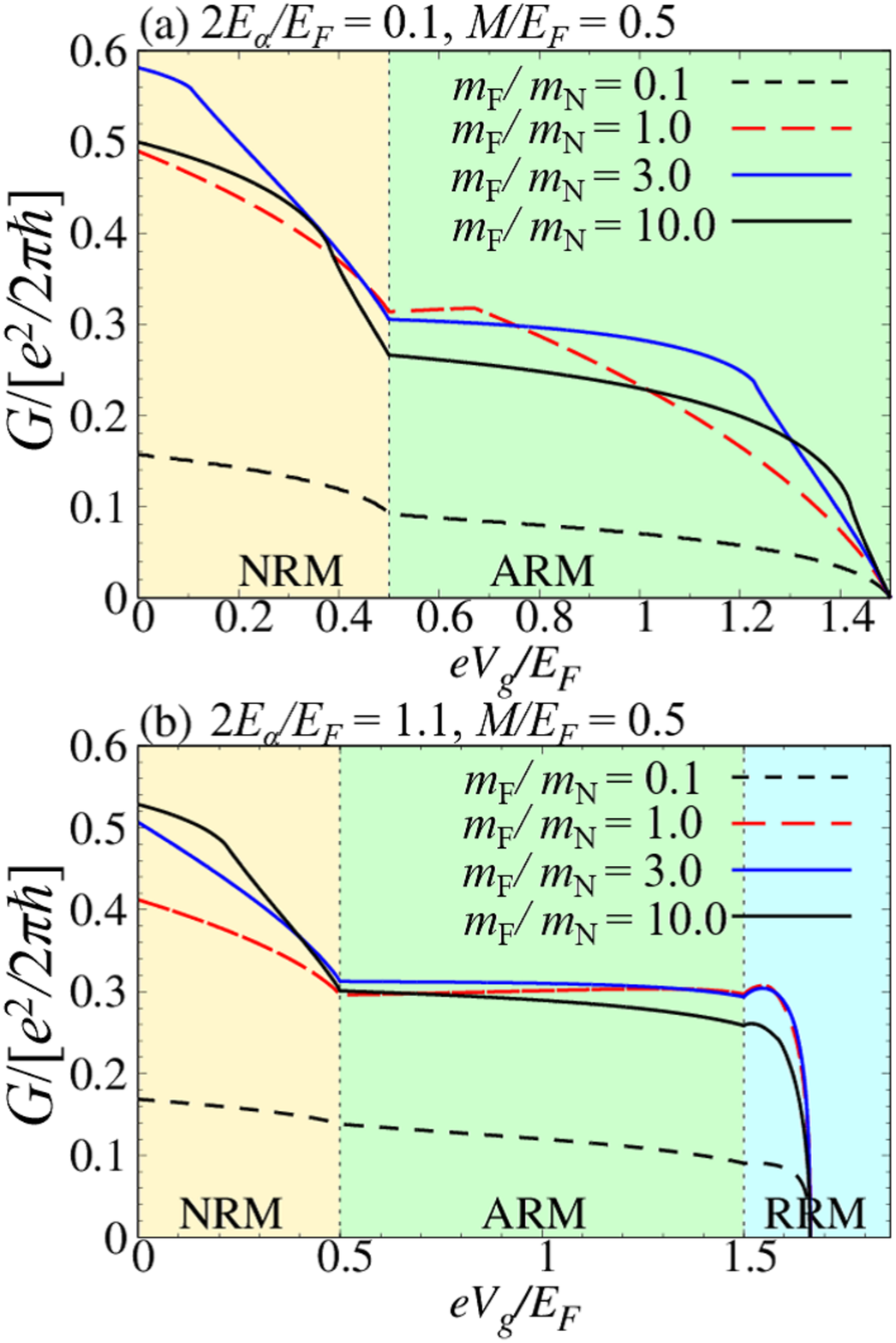}
\end{center}
\caption{(Color Online) $V_g$ dependence of $G$ for several $m_{\rm F}/m_{\rm N}$ values in (a) weak RSOC ($2E_\alpha /E_F=0.1$) and (b) strong RSOC ($2E_\alpha /E_F=1.1$).
(a) and (b) correspond to the band structure of Fig. \ref{structure}(c) and \ref{structure}(b), respectively.
Here, we set $Uk_F/E_F=0.0$.}
\label{NMFRMG_m}
\end{figure}
%Figure end

Next, we change the ratio of the effective mass in the each side.
Figures \ref{NMFRMG_m}(a) and \ref{NMFRMG_m}(b) show the $V_g$ dependence of $G$ for any ratio of the effective mass, in the presence of weak RSOC ($2E_\alpha <M: 2E_\alpha /E_F=0.1, M/E_F=0.5$) and in strong RSOC ($2E_\alpha >M: 2E_\alpha /E_F=1.1, M/E_F=0.5$).
Here, we set $Uk_F/E_F=0$.
In Fig. \ref{NMFRMG_m}(a)  ($m_{\rm F}/m_{\rm N}=3.0, 10.0$), in ARM, the $V_g$ dependence of $G$ is similar to that of $G_{\rm 1D}$ when the Fermi level is far from the band bottom, and it is the same as that of $G_{\rm 2D}$ when the Fermi level is near the band bottom.
For other results in the Figs.  \ref{NMFRMG_m}(a) and \ref{NMFRMG_m}(b), the magnitude of $G$ depends on $m_{\rm F}/m_{\rm N}$,
but the $V_g$ dependence is almost qualitatively the same for 
the results in Fig. \ref{NMFRMG_a}.

Thus, by tuning the Fermi level, the $V_g$ dependence of the conductance is dramatically changed.
In some cases, in spite of the 2D junction, the $V_g$ dependence is similar to that in conventional 1D system.
In particular, in ARM, which are caused by both RSOC and magnetization, the $V_g$ dependence of $G$ is almost determined on the relation between $E_\alpha$ and $M$.
This is the main result in this work.

%------------------------------------------
%Discussion
%------------------------------------------
\section{Discussion}\label{sec:IV}
We calculate the DOS in FRM with the gate voltage by using Green's functions. 
Details of the calculation are shown in the Appendix A.
The particle DOS at the Fermi level in the FRM, DOS$(E_F-eV_g)$, is analytically given by 
\eq{
{\rm DOS}(E_F-eV_g)=\left \{
\begin{aligned}
2\nu_e\qquad&(\rm{NRM}),\\
\nu_e \sqrt{\frac{E_\alpha}{\epsilon_F}}+\nu_e\qquad &(\rm{ARM}), \\
2\nu_e \sqrt{\frac{E_\alpha}{\epsilon_F}}\qquad &(\rm{RRM}).
\end{aligned}
\right.
\label{particleDOS}
}
where $2\nu_e=m_{\rm F}/(\pi\hbar^2)$ is the DOS in 2D system in the absence of RSOC and magnetization, and $\epsilon_F=E_F-eV_g+E_\alpha+E_c$ is a linear function of $E_F-eV_g$.
%
%Figure
\begin{figure}[t]
\begin{center}
\includegraphics[width=85mm]{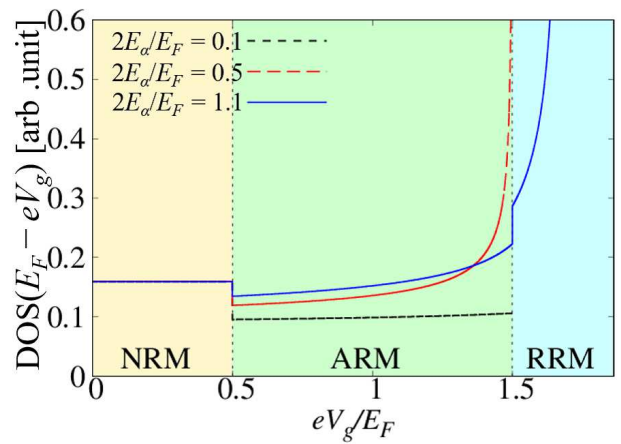}
\end{center}
\caption{(Color Online) $V_g$ dependence of the DOS in the 2D FRM for several $E_\alpha$ values at $M/E_F=0.5$.
The Fermi energy $E_F$ is shifted by the gate voltage $V_g$.
$V_g$ corresponds to the position of the Fermi level in FRM. 
The DOS in NRM is independent of RSOC and is exactly equal to that in a 2DEG. 
The RSOC parameter $2E_\alpha /E_F <1.0$ and $2E_\alpha /E_F > 1.0$ corresponds to cases (iii) and (ii), respectively (See Fig. \ref{structure}).
In the weak RSOC ($2E_\alpha /E_F =0.1$), the DOS is almost independent of $eV_g$, namely, position of the Fermi level, even in the ARM. 
However, in a strong RSOC ($2E_\alpha \simeq M$), the DOS in ARM and RRM depends on $eV_g$.
In particular, the $V_g$ dependence of the DOS in the RRM is equal to that in 1D-NM in the absence of the RSOC and exchange coupling.}
\label{DOSNMFRM_a}
\end{figure} 
%Figure end
%
These results are plotted in Fig. \ref{DOSNMFRM_a} as a function of $V_g$ for several $E_\alpha$ values at $M/E_F=0.5$.
From these equations, 
we find that in NRM ($eV_g/E_F<0.5$), the conventional DOS, which is independent of the Fermi level, is obtained.
In addition, the RSOC and magnetization do not affect the DOS in NRM. 
However, in ARM ($0.5<eV_g/E_F<1.5$) and RRM ($1.5< eV_g/E_F$), 
the DOS depends on the Fermi level, which can be caused by the RSOC and magnetization.
In particular, in RRM, the DOS is proportional to $1/\sqrt{\epsilon_F}$. 
Hence, the DOS in RRM is almost equivalent 
to that in a 1D electron gas (1DEG).
In ARM, the DOS is composed of two parts.
These parts can be regarded as DOS in a 2D electron gas (2DEG) and that in a 1DEG, as shown in Eq. (\ref{particleDOS}).
Furthermore, we find that in the weak RSOC ($2E_\alpha /E_F=0.1$) region, the DOS tends to behave like the DOS in a 2DEG. 
In a strong RSOC, the DOS is approximately equal to that in a 1DEG.
Thus, the energy dependence of the DOS is affected by $V_g$.

\begin{figure}[h]
\begin{center}
\includegraphics[width=85mm]{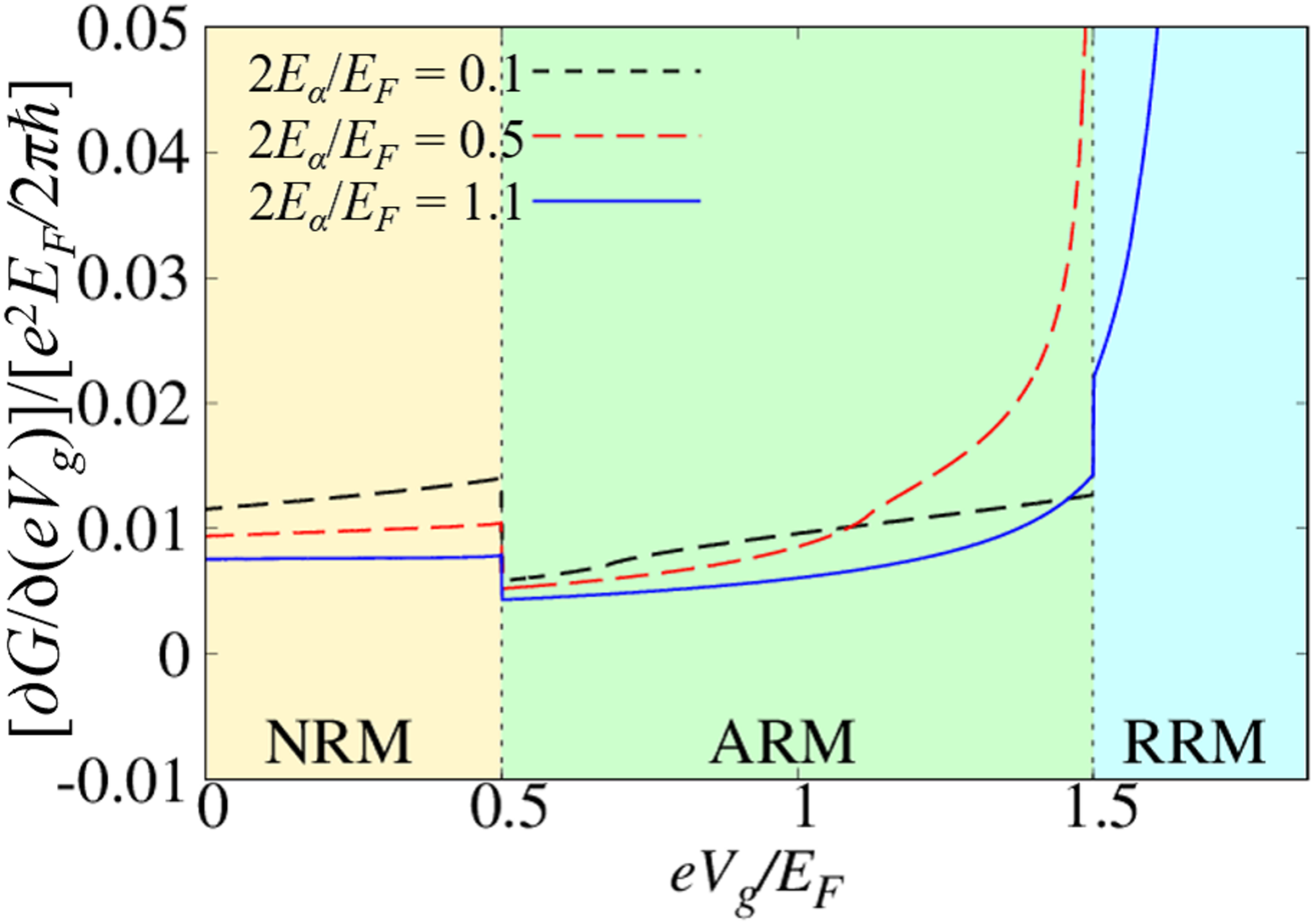}
\end{center}
\caption{(Color Online) $V_g$ dependence of  $-\partial G/\partial (eV_g)$ for several $E_\alpha$ values at $M/E_F=0.5$, $Uk_F/E_F=10.0$, and $m_{\rm F}/m_{\rm N}=1.0$.}
\label{dG}
\end{figure}

We find that $V_g$ dependence of $-\partial G/\partial (eV_g)$ is similar to that of the DOS in FRM (see Figs. \ref{DOSNMFRM_a} and \ref{dG}).
We confirmed that this tendency becomes more prominent when the magnitude of $U$ is sufficiently large.
%
%Figure
%Figure end
%
Therefore, we expect that the characteristic dimensionality of the DOS in FRM could directly reflect on the $V_g$ dependence of $G$ (see Table \ref{t2}). 
%Table
\begin{table}[H]
\centering
\caption{$V_g$ dependence of the tunneling conductance $G$ and DOS in NRM, ARM, and RRM.}
\begin{tabular}{cccc}
\hline
\qquad & NRM  & ARM & RRM \\
\hline
$G$ & 2D-like  & 2D-like (Weak RSOC)& 1D-like\\
                   &            &  1D-like (Strong RSOC) &          \\
DOS            & 2D        & 1D+2D & 1D\\
\hline
\end{tabular}
\label{t2}
\end{table}
%Table end

%------------------------------------------
%Conclusion
%------------------------------------------
\section{Conclusion}
\label{sec:V}
We theoretically studied tunneling conductance $G(V_g)$ in a 2D 
NM/FRM junction, where the Fermi level of the FRM is tuned by the applied gate voltage ($V_g$). 
This paper focuses on the nontrivial metallic states in FRM, $i.e.$, NRM, ARM, and RRM, which are realized in the presence of both RSOC and magnetization.
First, we find that ARM exhibits a unique behavior of the $V_g$ dependence of $G$, because of the coexistence of RSOC and exchange coupling $M$. 
The conductance strongly depends on the magnitude of the RSOC. 
For weak RSOC, the line shape of $G(V_g)$ is equal to the $V_g$ dependence of the conventional tunneling conductance in the absence of RSOC and magnetization.
However, in the presence of strong RSOC, 
$G(V_g)$ is almost independent of $V_g$. As a result, even in a 2D system, the $V_g$ dependence is almost equivalent to that in a 1D junction.
Second, we determined that the $V_g$ dependence of $G$ in RRM is similar to that in a conventional 1D junction.
Such an anomalous $V_g$ dependence of $G(V_g)$ could benefit from the characteristic dimensionality of the DOS of each of the nontrivial metallic states. 

For verifying the obtained results by experiments, thin heterostructure of Pt/Co/AlO$_x$ is promising since FRM can be realized\cite{Miron10}.
The RSOC can be caused by the inversion symmetry breaking along the layered direction, and the exchange coupling is due to the Co film.
By using the RSOC coefficient ($\alpha\approx 1$ eV\AA) and $\hbar^2/(2m_{\rm F})\approx 1$ ${\rm eV\AA^2}$, the Rashba energy is estimated as $E_\alpha\approx 0.1$ eV.
The magnitude of the exchange coupling could be estimated 
from the typical exchange coupling of ferromagnets, $M\sim 1$ eV, because the magnetization of the thin heterostructure is similar to that of its bulk \cite{Miron10}. 
Although its exchange coupling tends to be larger than the energy scale of the RSOC, the exchange coupling could be manipulated by an external magnetic field. 
The magnetic hysteresis of the magnetization is manipulated by using an applied magnetic field, and the exchange coupling is proportional to the magnetization.
Hence, the exchange coupling $M$ could be manipulated by an applied magnetic field.
Therefore, we expect that, in the heterojunction with a magnetic field applied along the out-of-plane direction, 
a situation of $2E_\alpha > M$ can be experimentally realized by using the property of magnetic hysteresis.
Moreover, we note that a thin-film heterojunction could be reasonable for substantially tuning the Fermi level using the gate voltage. 

Up to now, Rashba type spin-orbit coupling in 2D system has been studied in NRM, and several characteristic transport has been discovered in the field of spintronics.  
On the other hand, we systematically studied not only NRM but also ARM and RRM.
As a result, we theoretically discovered anomalous charge transport and DOS, which are also characteristic in the ARM and RRM (e.g., anomalous dimensionality of the DOS).
Therefore, unconventional charge and spin-related transport could be expected in the ARM and RRM.

In this paper, we have studied about charge transport properties of FRM. 
There are several remaining and future works. 
Since FRM has  specific spin structures with non-zero Berry curvature, 
we can expect new charge and spin transport phenomena such as unconventional Edelstein effect \cite{Taguchi17}. 
Besides this, to calculate tunneling magneto resistance (TMR) in FRM junctions is interesting.
TMR may show the different features depending on the gate voltage, 
where FRM is in the NRM, ARM or RRM regime. 
Secondly, to compare ARM and surface states of TI \cite{Hasan10} is an interesting topic.  
Both of these systems,  the electron's degree of freedom is reduced to be half due to the strong spin orbit coupling. 
By contrast to the surface state of TI \cite{Hasan10,Hasan11,Manchon15}, 
ARM does not have the time-reversal symmetry, and it has a kinetic term proportional to $k^2$.
At present, it is not clear how different physical properties appear between these two systems. 
Finally, the physical properties of RRM are not obvious, 
and exotic quantum phenomena might exist. 
We will study  these problems as a future work.

%------------------------------------------
%Acknowledgments
%------------------------------------------
%\begin{acknowledgment}
\section*{Acknowledgment}
We would like to thank K. Yada and K.T. Law for valuable discussion. 
This work was supported
by a Grant-in-Aid for Scientific Research on Innovative
Areas, Topological Material Science (Grants No. JP15H05851,
No. JP15H05853 No. JP15K21717),  a Grant-in-Aid
for Challenging Exploratory Research (Grant No. JP15K13498)
from the Ministry of Education, Culture, Sports, Science, and
Technology, Japan (MEXT), the Core Research for
Evolutional Science and Technology (CREST) of the Japan
Science and Technology Corporation (JST)(Grants No. JPMJCR14F1).
%\end{acknowledgment}

%------------------------------------------
%Appendix
%------------------------------------------
\appendix
\label{app1}
\section{Derivation of the density of states}
We show the detailed derivation of the DOS in FRM [see Eq. (\ref{particleDOS})].
The spin-dependent DOS is defined by
\eq{
\rho(E'_F)\equiv -\frac{1}{\pi}{\rm Im}\sum_{\bm{k}}\mathcal{G}_{\bm{k}}^{R}({\it E'_F}),
\label{rho}
}
where $E'_F=E_F-eV_g$ corresponds to the position of the Fermi surface.
$\mathcal{G}_{\bm{k}}^{R}({\it E'_F})\equiv{(E'_F-H_{FRM}+i\delta)}^{-1}$ is the retarded Green's function, and $\delta$ is an infinitesimal positive value.
From Eq. (\ref{rho}), we find that the off-diagonal components are zero ($\rho_{\uparrow\downarrow}=\rho_{\downarrow\uparrow}=0$) in every case.
The retarded Green's function can be divided into two parts as follows:
\eq{
\mathcal{G}_{\bm{k}}^{R}=\mathcal{G}_{\bm{k}}^{R+}+\mathcal{G}_{\bm{k}}^{R-},
}
where
\eq{
\mathcal{G}_{\bm{k}}^{R\pm}&=\frac{\Omega_\pm}{E'_F-E_\pm+i\delta},\\
\Omega_\pm&=\frac{1}{2}(1\pm \bm{n}\cdot\bm{\sigma}),\\
\bm{n}&\equiv-\frac{\alpha(\bm{k}\times\hat{\mathbf{z}})+M\hat{\mathbf{z}}}{\sqrt{\alpha^2 k^2+M^2}}.
}
Here, $\hat{\mathbf{z}}$ is a unit vector along {\it z}-axis.
$\mathcal{G}_{\bm{k}}^{R\pm}$ are the Green's function of $E_\pm$.
Therefore, Eq. (\ref{rho}) is decomposed as follows:
\eq{
\rho (E'_F)&=\rho^+(E'_F)+\rho^-(E'_F),\\
\rho^\pm(E'_F)&\equiv-\frac{1}{\pi}{\rm Im}\sum_{\bm{k}}\mathcal{G}_{\bm{k}}^{R\pm}({\it E'_F}).
}
We calculate $\rho^-(E'_F)$:
\eq{
\rho^-(E'_F)&=-\frac{1}{2\pi}{\rm Im}\sum_{\bm{k}}\frac{1-\bm{n}\cdot\bm{\sigma}}{E'_F-E_- +i\delta}\nonumber\\
             &=-\frac{1}{4\pi^2}{\rm Im}\int_0^\infty kdk\frac{1-n_z \sigma_z}{ E'_F-E_- +i\delta}\nonumber\\
            =\frac{\nu_e}{2\pi}{\rm Im}&\int_0^\infty  d\epsilon_0\frac{1- n_z(\epsilon_0)\sigma_z}{ \epsilon_0-2\sqrt{ E_\alpha}\sqrt{\epsilon_0+E_c}-({\it E'_F+i\delta})},\nonumber\\
}
where $\epsilon_0=\hbar^2k^2/(2m_{\rm F})$ is the kinetic energy in NM.
We set $\xi=\sqrt{\epsilon_0+E_c}$.
$\rho^-$ becomes
\eq{
\rho^-(E'_F)=\frac{\nu_e}{\pi}{\rm Im}\int_{\sqrt{E_c}}^\infty {\it d\xi}\frac{[1-n_z(\xi)\sigma_z]\xi}{\xi^2-2\sqrt{E_\alpha}\xi-(E'_F+E_c)-i\delta}.
}
$n_z$ is given by
\eq{
n_z=-\frac{M}{\sqrt{\alpha^2 k^2+M^2}}=-\sqrt{\frac{E_c}{\epsilon_0+E_c}}=-\frac{\sqrt{E_c}}{\xi}.
}
As a result, we can obtain
\eq{
\rho^-(E'_F)&=\frac{\nu_e}{\pi}{\rm Im}\int_{\sqrt{E_c}}^\infty  d\xi\frac{\xi+\sqrt{E_c}\sigma_z}{F_-(\xi)-i\delta}\nonumber\\
             &=\frac{\nu_e}{2i\pi}\int_{\sqrt{E_c}}^\infty d\xi\left[\frac{\xi+\sqrt{E_c}\sigma_z}{F_-(\xi)-i\delta}-\frac{\xi+\sqrt{E_c}\sigma_z}{F_-(\xi)+i\delta}\right]\nonumber\\
             &=\nu_e\int_{\sqrt{E_c}}^\infty d\xi\left(\xi+\sqrt{E_c}\sigma_z\right)\delta[F_-(\xi)].
\label{Fxi}
}
Here, $F_-(\xi)$ is expressed by
\eq{
F_-(\xi)=\xi^2-2\sqrt{E_\alpha}\xi-(E'_F+E_c)=(\xi-\xi_1)(\xi-\xi_2),
}
with $\xi_{1(2)}=\sqrt{E_\alpha}+(-)\sqrt{\epsilon_F}$, and $\epsilon_F=E_F-eV_g+E_\alpha+E_c$.
Then, $\delta[F_-(\xi)]$ in Eq. (\ref{Fxi}) is estimated by
\eq{
\delta[F_-(\xi)]&=\frac{1}{|\partial_{\xi}F_-(\xi_1)|} \delta(\xi-\xi_1) + \frac{1}{|\partial_{\xi}F_-(\xi_2)|}\delta(\xi-\xi_2)\nonumber\\
                &=\frac{1}{2\sqrt{\epsilon_F}} [\delta(\xi-\xi_1) + \delta(\xi-\xi_2)].
}
As a result, we have
\eq{
\rho^-(E'_F)&=\frac{\nu_e}{2\sqrt{\epsilon_F}}\int_{\sqrt{E_c}}^\infty d\xi\left(\xi+\sqrt{E_c}\sigma_z\right)\nonumber\\
             &\qquad\qquad\qquad\qquad\times[\delta(\xi-\xi_1) + \delta(\xi-\xi_2)]\nonumber\\
             &=\frac{\nu_e}{2}\left[\left(1+\frac{2E_\alpha+M\sigma_z}{2\sqrt{E_\alpha\epsilon_F}}\right)\theta(\sqrt{\epsilon_F}+\sqrt{E_\alpha}-\sqrt{E_c})\right.\nonumber\\
&\left.\qquad+\left(-1+\frac{2E_\alpha+M\sigma_z}{2\sqrt{E_\alpha\epsilon_F}}\right)\theta(-\sqrt{\epsilon_F}+\sqrt{E_\alpha}-\sqrt{E_c})\right].
\label{rho-}
}
To replace $\sqrt{E_\alpha}$ with $-\sqrt{E_\alpha}$, we obtain $\rho^+$ as follows: 
\eq{
\rho^+(E'_F)&=\frac{\nu_e}{2}\left(1-\frac{2E_\alpha+M\sigma_z}{2\sqrt{E_\alpha\epsilon_F}}\right)\theta(\sqrt{\epsilon_F}-\sqrt{E_\alpha}-\sqrt{E_c}).
\label{rho+}
}
From Eqs. (\ref{rho-}) and (\ref{rho+}), 
we can obtain the spin-dependent DOS $\rho(E'_F)$ as follows:
\eq{
\rho(E'_F)&=\frac{\nu_e}{2}\left[\left(1-\frac{2E_\alpha+M\sigma_z}{2\sqrt{E_\alpha\epsilon_F}}\right)\theta(\sqrt{\epsilon_F}-\sqrt{E_\alpha}-\sqrt{E_c})\right.\nonumber\\
&\qquad+\left(1+\frac{2E_\alpha+M\sigma_z}{2\sqrt{E_\alpha\epsilon_F}}\right)\theta(\sqrt{\epsilon_F}+\sqrt{E_\alpha}-\sqrt{E_c})\nonumber\\
&\left.\qquad+\left(-1+\frac{2E_\alpha+M\sigma_z}{2\sqrt{E_\alpha\epsilon_F}}\right)\theta(-\sqrt{\epsilon_F}+\sqrt{E_\alpha}-\sqrt{E_c})\right].
}
Here, the resulting DOS as Eq. (\ref{particleDOS}) is given by $\rm{Tr}[\rho]=\rho_{\uparrow\uparrow}+\rho_{\downarrow\downarrow}$ in case (ii).
The DOS in cases (i)-(iii) are shown in Table \ref{tableDOS}.
We can demonstrate the DOS for $M=0$ agrees with that in Ref. 24.

%Table
\begin{table}[H]
\centering
\caption{$V_g$ dependence of the DOS in the Fermi surface of FRM for NRM, ARM, and RRM in cases (i)-(iii).
The position of the Fermi level becomes lower with increasing $V_g$.
Here, $\epsilon_F=E_F-eV_g+E_\alpha+E_c$ is a linear function of $E_F-eV_g$, and $2\nu_e$ is the DOS in 2DEG.
NRM, ARM, and RRM is realized 
in $0<eV_g<E_F-M$, $E_F-M<eV_g<E_F+M$, and $E_F+M<eV_g<E_F+E_\alpha+E_c$, respectively. }
\begin{tabular}{lccc}
\hline
\qquad Case& NRM  & ARM & RRM \\
\hline
(i)\qquad$M=0 $ & $2\nu_e$ & -&$2\nu_e\sqrt{\frac{E_\alpha}{\epsilon_F}}$\\
(ii)$0<M<2E_\alpha$   & $2\nu_e$ & $\nu_e \left[ 1+ \sqrt{\frac{E_\alpha}{\epsilon_F}}\right]$ &$2\nu_e\sqrt{\frac{E_\alpha}{\epsilon_F}}$\\
(iii)\quad$M\geq 2E_\alpha$ & $2\nu_e$ & $\nu_e \left[ 1+ \sqrt{\frac{E_\alpha}{\epsilon_F}}\right]$ &-\\
\hline
\end{tabular}
\label{tableDOS}
\end{table}
%Table end



\begin{thebibliography}{40}%
\makeatletter
\providecommand \@ifxundefined [1]{%
 \@ifx{#1\undefined}
}%
\providecommand \@ifnum [1]{%
 \ifnum #1\expandafter \@firstoftwo
 \else \expandafter \@secondoftwo
 \fi
}%
\providecommand \@ifx [1]{%
 \ifx #1\expandafter \@firstoftwo
 \else \expandafter \@secondoftwo
 \fi
}%
\providecommand \natexlab [1]{#1}%
\providecommand \enquote  [1]{``#1''}%
\providecommand \bibnamefont  [1]{#1}%
\providecommand \bibfnamefont [1]{#1}%
\providecommand \citenamefont [1]{#1}%
\providecommand \href@noop [0]{\@secondoftwo}%
\providecommand \href [0]{\begingroup \@sanitize@url \@href}%
\providecommand \@href[1]{\@@startlink{#1}\@@href}%
\providecommand \@@href[1]{\endgroup#1\@@endlink}%
\providecommand \@sanitize@url [0]{\catcode `\\12\catcode `\$12\catcode
  `\&12\catcode `\#12\catcode `\^12\catcode `\_12\catcode `\%12\relax}%
\providecommand \@@startlink[1]{}%
\providecommand \@@endlink[0]{}%
\providecommand \url  [0]{\begingroup\@sanitize@url \@url }%
\providecommand \@url [1]{\endgroup\@href {#1}{\urlprefix }}%
\providecommand \urlprefix  [0]{URL }%
\providecommand \Eprint [0]{\href }%
\providecommand \doibase [0]{http://dx.doi.org/}%
\providecommand \selectlanguage [0]{\@gobble}%
\providecommand \bibinfo  [0]{\@secondoftwo}%
\providecommand \bibfield  [0]{\@secondoftwo}%
\providecommand \translation [1]{[#1]}%
\providecommand \BibitemOpen [0]{}%
\providecommand \bibitemStop [0]{}%
\providecommand \bibitemNoStop [0]{.\EOS\space}%
\providecommand \EOS [0]{\spacefactor3000\relax}%
\providecommand \BibitemShut  [1]{\csname bibitem#1\endcsname}%
\let\auto@bib@innerbib\@empty
%</preamble>
\bibitem [{\citenamefont {Yokoyama}\ \emph {et~al.}(2010)\citenamefont
  {Yokoyama}, \citenamefont {Tanaka},\ and\ \citenamefont
  {Nagaosa}}]{Yokoyama10}%
  \BibitemOpen
  \bibfield  {author} {\bibinfo {author} {\bibfnamefont {T.}~\bibnamefont
  {Yokoyama}}, \bibinfo {author} {\bibfnamefont {Y.}~\bibnamefont {Tanaka}}, \
  and\ \bibinfo {author} {\bibfnamefont {N.}~\bibnamefont {Nagaosa}},\
  }\href@noop {} {\bibfield  {journal} {\bibinfo  {journal} {Phys. Rev. B}\
  }\textbf {\bibinfo {volume} {81}},\ \bibinfo {pages} {121401} (\bibinfo
  {year} {2010})}\BibitemShut {NoStop}%
\bibitem [{\citenamefont {Mondal}\ \emph {et~al.}(2010)\citenamefont {Mondal},
  \citenamefont {Sen}, \citenamefont {Sengupta},\ and\ \citenamefont
  {Shankar}}]{Mondal10}%
  \BibitemOpen
  \bibfield  {author} {\bibinfo {author} {\bibfnamefont {S.}~\bibnamefont
  {Mondal}}, \bibinfo {author} {\bibfnamefont {D.}~\bibnamefont {Sen}},
  \bibinfo {author} {\bibfnamefont {K.}~\bibnamefont {Sengupta}}, \ and\
  \bibinfo {author} {\bibfnamefont {R.}~\bibnamefont {Shankar}},\ }\href@noop
  {} {\bibfield  {journal} {\bibinfo  {journal} {Phys. Rev. Lett.}\ }\textbf
  {\bibinfo {volume} {104}},\ \bibinfo {pages} {046403} (\bibinfo {year}
  {2010})}\BibitemShut {NoStop}%
\bibitem [{\citenamefont {Taguchi}\ \emph {et~al.}(2014)\citenamefont
  {Taguchi}, \citenamefont {Yokoyama},\ and\ \citenamefont
  {Tanaka}}]{Taguchi14}%
  \BibitemOpen
  \bibfield  {author} {\bibinfo {author} {\bibfnamefont {K.}~\bibnamefont
  {Taguchi}}, \bibinfo {author} {\bibfnamefont {T.}~\bibnamefont {Yokoyama}}, \
  and\ \bibinfo {author} {\bibfnamefont {Y.}~\bibnamefont {Tanaka}},\
  }\href@noop {} {\bibfield  {journal} {\bibinfo  {journal} {Phys. Rev. B}\
  }\textbf {\bibinfo {volume} {89}},\ \bibinfo {pages} {085407} (\bibinfo
  {year} {2014})}\BibitemShut {NoStop}%
\bibitem [{\citenamefont {Molenkamp}\ \emph {et~al.}(2001)\citenamefont
  {Molenkamp}, \citenamefont {Schmidt},\ and\ \citenamefont
  {Bauer}}]{Molenkamp01}%
  \BibitemOpen
  \bibfield  {author} {\bibinfo {author} {\bibfnamefont {L.~W.}\ \bibnamefont
  {Molenkamp}}, \bibinfo {author} {\bibfnamefont {G.}~\bibnamefont {Schmidt}},
  \ and\ \bibinfo {author} {\bibfnamefont {G.~E.}\ \bibnamefont {Bauer}},\
  }\href@noop {} {\bibfield  {journal} {\bibinfo  {journal} {Phys. Rev. B}\
  }\textbf {\bibinfo {volume} {64}},\ \bibinfo {pages} {121202} (\bibinfo
  {year} {2001})}\BibitemShut {NoStop}%
\bibitem [{\citenamefont {Matsuyama}\ \emph {et~al.}(2002)\citenamefont
  {Matsuyama}, \citenamefont {Hu}, \citenamefont {Grundler}, \citenamefont
  {Meier},\ and\ \citenamefont {Merkt}}]{Matsuyama02}%
  \BibitemOpen
  \bibfield  {author} {\bibinfo {author} {\bibfnamefont {T.}~\bibnamefont
  {Matsuyama}}, \bibinfo {author} {\bibfnamefont {C.-M.}\ \bibnamefont {Hu}},
  \bibinfo {author} {\bibfnamefont {D.}~\bibnamefont {Grundler}}, \bibinfo
  {author} {\bibfnamefont {G.}~\bibnamefont {Meier}}, \ and\ \bibinfo {author}
  {\bibfnamefont {U.}~\bibnamefont {Merkt}},\ }\href@noop {} {\bibfield
  {journal} {\bibinfo  {journal} {Phys. Rev. B}\ }\textbf {\bibinfo {volume}
  {65}},\ \bibinfo {pages} {155322} (\bibinfo {year} {2002})}\BibitemShut
  {NoStop}%
\bibitem [{\citenamefont {Jiang}\ and\ \citenamefont {Jalil}(2003)}]{Jiang03}%
  \BibitemOpen
  \bibfield  {author} {\bibinfo {author} {\bibfnamefont {Y.}~\bibnamefont
  {Jiang}}\ and\ \bibinfo {author} {\bibfnamefont {M.}~\bibnamefont {Jalil}},\
  }\href@noop {} {\bibfield  {journal} {\bibinfo  {journal} {J. Phys: Cond.
  Mat.}\ }\textbf {\bibinfo {volume} {15}},\ \bibinfo {pages} {L31} (\bibinfo
  {year} {2003})}\BibitemShut {NoStop}%
\bibitem [{\citenamefont {Ramaglia}\ \emph {et~al.}(2003)\citenamefont
  {Ramaglia}, \citenamefont {Bercioux}, \citenamefont {Cataudella},
  \citenamefont {Filippis}, \citenamefont {Perroni},\ and\ \citenamefont
  {Ventriglia}}]{Ramaglia03}%
  \BibitemOpen
  \bibfield  {author} {\bibinfo {author} {\bibfnamefont {V.~M.}\ \bibnamefont
  {Ramaglia}}, \bibinfo {author} {\bibfnamefont {D.}~\bibnamefont {Bercioux}},
  \bibinfo {author} {\bibfnamefont {V.}~\bibnamefont {Cataudella}}, \bibinfo
  {author} {\bibfnamefont {G.~D.}\ \bibnamefont {Filippis}}, \bibinfo {author}
  {\bibfnamefont {C.}~\bibnamefont {Perroni}}, \ and\ \bibinfo {author}
  {\bibfnamefont {F.}~\bibnamefont {Ventriglia}},\ }\href@noop {} {\bibfield
  {journal} {\bibinfo  {journal} {Eur. Phys. J. B-Condensed Matter and Complex
  Systems}\ }\textbf {\bibinfo {volume} {36}},\ \bibinfo {pages} {365}
  (\bibinfo {year} {2003})}\BibitemShut {NoStop}%
\bibitem [{\citenamefont {Yokoyama}\ \emph {et~al.}(2006)\citenamefont
  {Yokoyama}, \citenamefont {Tanaka},\ and\ \citenamefont
  {Inoue}}]{Yokoyama06}%
  \BibitemOpen
  \bibfield  {author} {\bibinfo {author} {\bibfnamefont {T.}~\bibnamefont
  {Yokoyama}}, \bibinfo {author} {\bibfnamefont {Y.}~\bibnamefont {Tanaka}}, \
  and\ \bibinfo {author} {\bibfnamefont {J.}~\bibnamefont {Inoue}},\
  }\href@noop {} {\bibfield  {journal} {\bibinfo  {journal} {Phys. Rev. B}\
  }\textbf {\bibinfo {volume} {74}},\ \bibinfo {pages} {035318} (\bibinfo
  {year} {2006})}\BibitemShut {NoStop}%
\bibitem [{\citenamefont {Srisongmuang}\ \emph {et~al.}(2008)\citenamefont
  {Srisongmuang}, \citenamefont {Pairor},\ and\ \citenamefont
  {Berciu}}]{Srisongmuang08}%
  \BibitemOpen
  \bibfield  {author} {\bibinfo {author} {\bibfnamefont {B.}~\bibnamefont
  {Srisongmuang}}, \bibinfo {author} {\bibfnamefont {P.}~\bibnamefont
  {Pairor}}, \ and\ \bibinfo {author} {\bibfnamefont {M.}~\bibnamefont
  {Berciu}},\ }\href@noop {} {\bibfield  {journal} {\bibinfo  {journal} {Phys.
  Rev. B}\ }\textbf {\bibinfo {volume} {78}},\ \bibinfo {pages} {155317}
  (\bibinfo {year} {2008})}\BibitemShut {NoStop}%
\bibitem [{\citenamefont {Matos-Abiague}\ and\ \citenamefont
  {Fabian}(2009)}]{Matos-Abiague09}%
  \BibitemOpen
  \bibfield  {author} {\bibinfo {author} {\bibfnamefont {A.}~\bibnamefont
  {Matos-Abiague}}\ and\ \bibinfo {author} {\bibfnamefont {J.}~\bibnamefont
  {Fabian}},\ }\href@noop {} {\bibfield  {journal} {\bibinfo  {journal} {Phys.
  Rev. B}\ }\textbf {\bibinfo {volume} {79}},\ \bibinfo {pages} {155303}
  (\bibinfo {year} {2009})}\BibitemShut {NoStop}%
\bibitem [{\citenamefont {Zhang}\ and\ \citenamefont {Xu}(2013)}]{Zhang13}%
  \BibitemOpen
  \bibfield  {author} {\bibinfo {author} {\bibfnamefont {P.}~\bibnamefont
  {Zhang}}\ and\ \bibinfo {author} {\bibfnamefont {M.}~\bibnamefont {Xu}},\
  }\href@noop {} {\bibfield  {journal} {\bibinfo  {journal} {Sci. China Phys.,
  Mechanics and Astronomy}\ }\textbf {\bibinfo {volume} {56}},\ \bibinfo
  {pages} {1514} (\bibinfo {year} {2013})}\BibitemShut {NoStop}%
\bibitem [{\citenamefont {Jantayod}\ and\ \citenamefont
  {Pairor}(2013)}]{Jantayod13}%
  \BibitemOpen
  \bibfield  {author} {\bibinfo {author} {\bibfnamefont {A.}~\bibnamefont
  {Jantayod}}\ and\ \bibinfo {author} {\bibfnamefont {P.}~\bibnamefont
  {Pairor}},\ }\href@noop {} {\bibfield  {journal} {\bibinfo  {journal} {Phys.
  E: Low-dimensional Systems and Nanostructures}\ }\textbf {\bibinfo {volume}
  {48}},\ \bibinfo {pages} {111} (\bibinfo {year} {2013})}\BibitemShut
  {NoStop}%
\bibitem [{\citenamefont {Jantayod}\ and\ \citenamefont
  {Pairor}(2015)}]{Jantayod15}%
  \BibitemOpen
  \bibfield  {author} {\bibinfo {author} {\bibfnamefont {A.}~\bibnamefont
  {Jantayod}}\ and\ \bibinfo {author} {\bibfnamefont {P.}~\bibnamefont
  {Pairor}},\ }\href@noop {} {\bibfield  {journal} {\bibinfo  {journal}
  {Superlattices and Microstructures}\ }\textbf {\bibinfo {volume} {88}},\
  \bibinfo {pages} {541} (\bibinfo {year} {2015})}\BibitemShut {NoStop}%
\bibitem [{\citenamefont {Grundler}(2001)}]{Grundler01}%
  \BibitemOpen
  \bibfield  {author} {\bibinfo {author} {\bibfnamefont {D.}~\bibnamefont
  {Grundler}},\ }\href@noop {} {\bibfield  {journal} {\bibinfo  {journal}
  {Phys. Rev. B}\ }\textbf {\bibinfo {volume} {63}},\ \bibinfo {pages} {161307}
  (\bibinfo {year} {2001})}\BibitemShut {NoStop}%
\bibitem [{\citenamefont {Larsen}\ \emph {et~al.}(2002)\citenamefont {Larsen},
  \citenamefont {Lunde},\ and\ \citenamefont {Flensberg}}]{Larsen02}%
  \BibitemOpen
  \bibfield  {author} {\bibinfo {author} {\bibfnamefont {M.~H.}\ \bibnamefont
  {Larsen}}, \bibinfo {author} {\bibfnamefont {A.~M.}\ \bibnamefont {Lunde}}, \
  and\ \bibinfo {author} {\bibfnamefont {K.}~\bibnamefont {Flensberg}},\
  }\href@noop {} {\bibfield  {journal} {\bibinfo  {journal} {Phys. Rev. B}\
  }\textbf {\bibinfo {volume} {66}},\ \bibinfo {pages} {033304} (\bibinfo
  {year} {2002})}\BibitemShut {NoStop}%
\bibitem [{\citenamefont {St{\v{r}}eda}\ and\ \citenamefont
  {{\v{S}}eba}(2003)}]{Streda03}%
  \BibitemOpen
  \bibfield  {author} {\bibinfo {author} {\bibfnamefont {P.}~\bibnamefont
  {St{\v{r}}eda}}\ and\ \bibinfo {author} {\bibfnamefont {P.}~\bibnamefont
  {{\v{S}}eba}},\ }\href@noop {} {\bibfield  {journal} {\bibinfo  {journal}
  {Phys. Rev. Lett.}\ }\textbf {\bibinfo {volume} {90}},\ \bibinfo {pages}
  {256601} (\bibinfo {year} {2003})}\BibitemShut {NoStop}%
\bibitem [{\citenamefont {Krupin}\ \emph {et~al.}(2005)\citenamefont {Krupin},
  \citenamefont {Bihlmayer}, \citenamefont {Starke}, \citenamefont {Gorovikov},
  \citenamefont {Prieto}, \citenamefont {D{\"o}brich}, \citenamefont
  {Bl{\"u}gel},\ and\ \citenamefont {Kaindl}}]{Krupin05}%
  \BibitemOpen
  \bibfield  {author} {\bibinfo {author} {\bibfnamefont {O.}~\bibnamefont
  {Krupin}}, \bibinfo {author} {\bibfnamefont {G.}~\bibnamefont {Bihlmayer}},
  \bibinfo {author} {\bibfnamefont {K.}~\bibnamefont {Starke}}, \bibinfo
  {author} {\bibfnamefont {S.}~\bibnamefont {Gorovikov}}, \bibinfo {author}
  {\bibfnamefont {J.}~\bibnamefont {Prieto}}, \bibinfo {author} {\bibfnamefont
  {K.}~\bibnamefont {D{\"o}brich}}, \bibinfo {author} {\bibfnamefont
  {S.}~\bibnamefont {Bl{\"u}gel}}, \ and\ \bibinfo {author} {\bibfnamefont
  {G.}~\bibnamefont {Kaindl}},\ }\href@noop {} {\bibfield  {journal} {\bibinfo
  {journal} {Phys. Rev. B}\ }\textbf {\bibinfo {volume} {71}},\ \bibinfo
  {pages} {201403} (\bibinfo {year} {2005})}\BibitemShut {NoStop}%
\bibitem [{\citenamefont {S{\'a}nchez}\ \emph {et~al.}(2008)\citenamefont
  {S{\'a}nchez}, \citenamefont {Serra},\ and\ \citenamefont
  {Choi}}]{Sanchez08}%
  \BibitemOpen
  \bibfield  {author} {\bibinfo {author} {\bibfnamefont {D.}~\bibnamefont
  {S{\'a}nchez}}, \bibinfo {author} {\bibfnamefont {L.}~\bibnamefont {Serra}},
  \ and\ \bibinfo {author} {\bibfnamefont {M.-S.}\ \bibnamefont {Choi}},\
  }\href@noop {} {\bibfield  {journal} {\bibinfo  {journal} {Phys. Rev. B}\
  }\textbf {\bibinfo {volume} {77}},\ \bibinfo {pages} {035315} (\bibinfo
  {year} {2008})}\BibitemShut {NoStop}%
\bibitem [{\citenamefont {Fallahi}\ and\ \citenamefont
  {Ghanaatshoar}(2012)}]{Fallahi12}%
  \BibitemOpen
  \bibfield  {author} {\bibinfo {author} {\bibfnamefont {V.}~\bibnamefont
  {Fallahi}}\ and\ \bibinfo {author} {\bibfnamefont {M.}~\bibnamefont
  {Ghanaatshoar}},\ }\href@noop {} {\bibfield  {journal} {\bibinfo  {journal}
  {Phys. Status Solidi (B)}\ }\textbf {\bibinfo {volume} {249}},\ \bibinfo
  {pages} {1077} (\bibinfo {year} {2012})}\BibitemShut {NoStop}%
\bibitem [{\citenamefont {Pang}\ and\ \citenamefont {Wang}(2012)}]{Pang12}%
  \BibitemOpen
  \bibfield  {author} {\bibinfo {author} {\bibfnamefont {M.}~\bibnamefont
  {Pang}}\ and\ \bibinfo {author} {\bibfnamefont {C.}~\bibnamefont {Wang}},\
  }\href@noop {} {\bibfield  {journal} {\bibinfo  {journal} {Phys. E:
  Low-dimensional Systems and Nanostructures}\ }\textbf {\bibinfo {volume}
  {44}},\ \bibinfo {pages} {1636} (\bibinfo {year} {2012})}\BibitemShut
  {NoStop}%
\bibitem [{\citenamefont {Tang}\ \emph {et~al.}(2012)\citenamefont {Tang},
  \citenamefont {Chang},\ and\ \citenamefont {Cheng}}]{Tang12}%
  \BibitemOpen
  \bibfield  {author} {\bibinfo {author} {\bibfnamefont {C.-S.}\ \bibnamefont
  {Tang}}, \bibinfo {author} {\bibfnamefont {S.-Y.}\ \bibnamefont {Chang}}, \
  and\ \bibinfo {author} {\bibfnamefont {S.-J.}\ \bibnamefont {Cheng}},\
  }\href@noop {} {\bibfield  {journal} {\bibinfo  {journal} {Phys. Rev. B}\
  }\textbf {\bibinfo {volume} {86}},\ \bibinfo {pages} {125321} (\bibinfo
  {year} {2012})}\BibitemShut {NoStop}%
\bibitem [{\citenamefont {Tang}\ \emph {et~al.}(2016)\citenamefont {Tang},
  \citenamefont {Keng}, \citenamefont {Abdullah},\ and\ \citenamefont
  {Gudmundsson}}]{Tang16}%
  \BibitemOpen
  \bibfield  {author} {\bibinfo {author} {\bibfnamefont {C.-S.}\ \bibnamefont
  {Tang}}, \bibinfo {author} {\bibfnamefont {J.-A.}\ \bibnamefont {Keng}},
  \bibinfo {author} {\bibfnamefont {N.~R.}\ \bibnamefont {Abdullah}}, \ and\
  \bibinfo {author} {\bibfnamefont {V.}~\bibnamefont {Gudmundsson}},\
  }\href@noop {} {\bibfield  {journal} {\bibinfo  {journal} {arXiv:1612.05899}\
  } (\bibinfo {year} {2016})}\BibitemShut {NoStop}%
\bibitem [{\citenamefont {Fukumoto}\ \emph {et~al.}(2015)\citenamefont
  {Fukumoto}, \citenamefont {Taguchi}, \citenamefont {Kobayashi},\ and\
  \citenamefont {Tanaka}}]{Fukumoto15}%
  \BibitemOpen
  \bibfield  {author} {\bibinfo {author} {\bibfnamefont {T.}~\bibnamefont
  {Fukumoto}}, \bibinfo {author} {\bibfnamefont {K.}~\bibnamefont {Taguchi}},
  \bibinfo {author} {\bibfnamefont {S.}~\bibnamefont {Kobayashi}}, \ and\
  \bibinfo {author} {\bibfnamefont {Y.}~\bibnamefont {Tanaka}},\ }\href@noop {}
  {\bibfield  {journal} {\bibinfo  {journal} {Phys. Rev. B}\ }\textbf {\bibinfo
  {volume} {92}},\ \bibinfo {pages} {144514} (\bibinfo {year}
  {2015})}\BibitemShut {NoStop}%
\bibitem [{\citenamefont {Han}\ \emph {et~al.}(2015)\citenamefont {Han},
  \citenamefont {Serra},\ and\ \citenamefont {Choi}}]{Han15}%
  \BibitemOpen
  \bibfield  {author} {\bibinfo {author} {\bibfnamefont {S.}~\bibnamefont
  {Han}}, \bibinfo {author} {\bibfnamefont {L.}~\bibnamefont {Serra}}, \ and\
  \bibinfo {author} {\bibfnamefont {M.-S.}\ \bibnamefont {Choi}},\ }\href@noop
  {} {\bibfield  {journal} {\bibinfo  {journal} {J. Phys: Cond. Mat.}\ }\textbf
  {\bibinfo {volume} {27}},\ \bibinfo {pages} {255002} (\bibinfo {year}
  {2015})}\BibitemShut {NoStop}%
\bibitem [{\citenamefont {W{\'o}jcik}\ \emph {et~al.}(2015)\citenamefont
  {W{\'o}jcik}, \citenamefont {Adamowski}, \citenamefont {Wo{\l}oszyn},\ and\
  \citenamefont {Spisak}}]{Wojcik15}%
  \BibitemOpen
  \bibfield  {author} {\bibinfo {author} {\bibfnamefont {P.}~\bibnamefont
  {W{\'o}jcik}}, \bibinfo {author} {\bibfnamefont {J.}~\bibnamefont
  {Adamowski}}, \bibinfo {author} {\bibfnamefont {M.}~\bibnamefont
  {Wo{\l}oszyn}}, \ and\ \bibinfo {author} {\bibfnamefont {B.}~\bibnamefont
  {Spisak}},\ }\href@noop {} {\bibfield  {journal} {\bibinfo  {journal} {J.
  Appl. Phys.}\ }\textbf {\bibinfo {volume} {118}},\ \bibinfo {pages} {014302}
  (\bibinfo {year} {2015})}\BibitemShut {NoStop}%
\bibitem [{\citenamefont {Miron}\ \emph {et~al.}(2010)\citenamefont {Miron},
  \citenamefont {Gaudin}, \citenamefont {Auffret}, \citenamefont {Rodmacq},
  \citenamefont {Schuhl}, \citenamefont {Pizzini}, \citenamefont {Vogel},\ and\
  \citenamefont {Gambardella}}]{Miron10}%
  \BibitemOpen
  \bibfield  {author} {\bibinfo {author} {\bibfnamefont {I.~M.}\ \bibnamefont
  {Miron}}, \bibinfo {author} {\bibfnamefont {G.}~\bibnamefont {Gaudin}},
  \bibinfo {author} {\bibfnamefont {S.}~\bibnamefont {Auffret}}, \bibinfo
  {author} {\bibfnamefont {B.}~\bibnamefont {Rodmacq}}, \bibinfo {author}
  {\bibfnamefont {A.}~\bibnamefont {Schuhl}}, \bibinfo {author} {\bibfnamefont
  {S.}~\bibnamefont {Pizzini}}, \bibinfo {author} {\bibfnamefont
  {J.}~\bibnamefont {Vogel}}, \ and\ \bibinfo {author} {\bibfnamefont
  {P.}~\bibnamefont {Gambardella}},\ }\href@noop {} {\bibfield  {journal}
  {\bibinfo  {journal} {Nat. Mater}\ }\textbf {\bibinfo {volume} {9}},\
  \bibinfo {pages} {230} (\bibinfo {year} {2010})}\BibitemShut {NoStop}%
\bibitem [{\citenamefont {Rashba}(1960)}]{Rashba60}%
  \BibitemOpen
  \bibfield  {author} {\bibinfo {author} {\bibfnamefont {E.~I.}\ \bibnamefont
  {Rashba}},\ }\href@noop {} {\bibfield  {journal} {\bibinfo  {journal} {Sov.
  Phys. Solid State}\ }\textbf {\bibinfo {volume} {2}},\ \bibinfo {pages}
  {1109} (\bibinfo {year} {1960})}\BibitemShut {NoStop}%
\bibitem [{\citenamefont {Cayao}\ \emph {et~al.}(2015)\citenamefont {Cayao},
  \citenamefont {Prada}, \citenamefont {San-Jose},\ and\ \citenamefont
  {Aguado}}]{Cayao15}%
  \BibitemOpen
  \bibfield  {author} {\bibinfo {author} {\bibfnamefont {J.}~\bibnamefont
  {Cayao}}, \bibinfo {author} {\bibfnamefont {E.}~\bibnamefont {Prada}},
  \bibinfo {author} {\bibfnamefont {P.}~\bibnamefont {San-Jose}}, \ and\
  \bibinfo {author} {\bibfnamefont {R.}~\bibnamefont {Aguado}},\ }\href@noop {}
  {\bibfield  {journal} {\bibinfo  {journal} {Phys. Rev. B}\ }\textbf {\bibinfo
  {volume} {91}},\ \bibinfo {pages} {024514} (\bibinfo {year}
  {2015})}\BibitemShut {NoStop}%
\bibitem [{\citenamefont {Ast}\ \emph {et~al.}(2007)\citenamefont {Ast},
  \citenamefont {Henk}, \citenamefont {Ernst}, \citenamefont {Moreschini},
  \citenamefont {Falub}, \citenamefont {Pacil{\'e}}, \citenamefont {Bruno},
  \citenamefont {Kern},\ and\ \citenamefont {Grioni}}]{Ast07}%
  \BibitemOpen
  \bibfield  {author} {\bibinfo {author} {\bibfnamefont {C.~R.}\ \bibnamefont
  {Ast}}, \bibinfo {author} {\bibfnamefont {J.}~\bibnamefont {Henk}}, \bibinfo
  {author} {\bibfnamefont {A.}~\bibnamefont {Ernst}}, \bibinfo {author}
  {\bibfnamefont {L.}~\bibnamefont {Moreschini}}, \bibinfo {author}
  {\bibfnamefont {M.~C.}\ \bibnamefont {Falub}}, \bibinfo {author}
  {\bibfnamefont {D.}~\bibnamefont {Pacil{\'e}}}, \bibinfo {author}
  {\bibfnamefont {P.}~\bibnamefont {Bruno}}, \bibinfo {author} {\bibfnamefont
  {K.}~\bibnamefont {Kern}}, \ and\ \bibinfo {author} {\bibfnamefont
  {M.}~\bibnamefont {Grioni}},\ }\href@noop {} {\bibfield  {journal} {\bibinfo
  {journal} {Phys. Rev. Lett.}\ }\textbf {\bibinfo {volume} {98}},\ \bibinfo
  {pages} {186807} (\bibinfo {year} {2007})}\BibitemShut {NoStop}%
\bibitem [{\citenamefont {Sablikov}\ and\ \citenamefont
  {Tkach}(2007)}]{Sablikov07}%
  \BibitemOpen
  \bibfield  {author} {\bibinfo {author} {\bibfnamefont {V.~A.}\ \bibnamefont
  {Sablikov}}\ and\ \bibinfo {author} {\bibfnamefont {Y.~Y.}\ \bibnamefont
  {Tkach}},\ }\href {\doibase 10.1103/PhysRevB.76.245321} {\bibfield  {journal}
  {\bibinfo  {journal} {Phys. Rev. B}\ }\textbf {\bibinfo {volume} {76}},\
  \bibinfo {pages} {245321} (\bibinfo {year} {2007})}\BibitemShut {NoStop}%
\bibitem [{\citenamefont {Ast}\ \emph {et~al.}(2008)\citenamefont {Ast},
  \citenamefont {Pacil{\'e}}, \citenamefont {Moreschini}, \citenamefont
  {Falub}, \citenamefont {Papagno}, \citenamefont {Kern}, \citenamefont
  {Grioni}, \citenamefont {Henk}, \citenamefont {Ernst}, \citenamefont
  {Ostanin} \emph {et~al.}}]{Ast08}%
  \BibitemOpen
  \bibfield  {author} {\bibinfo {author} {\bibfnamefont {C.~R.}\ \bibnamefont
  {Ast}}, \bibinfo {author} {\bibfnamefont {D.}~\bibnamefont {Pacil{\'e}}},
  \bibinfo {author} {\bibfnamefont {L.}~\bibnamefont {Moreschini}}, \bibinfo
  {author} {\bibfnamefont {M.~C.}\ \bibnamefont {Falub}}, \bibinfo {author}
  {\bibfnamefont {M.}~\bibnamefont {Papagno}}, \bibinfo {author} {\bibfnamefont
  {K.}~\bibnamefont {Kern}}, \bibinfo {author} {\bibfnamefont {M.}~\bibnamefont
  {Grioni}}, \bibinfo {author} {\bibfnamefont {J.}~\bibnamefont {Henk}},
  \bibinfo {author} {\bibfnamefont {A.}~\bibnamefont {Ernst}}, \bibinfo
  {author} {\bibfnamefont {S.}~\bibnamefont {Ostanin}},  \emph {et~al.},\
  }\href@noop {} {\bibfield  {journal} {\bibinfo  {journal} {Phys. Rev. B}\
  }\textbf {\bibinfo {volume} {77}},\ \bibinfo {pages} {081407} (\bibinfo
  {year} {2008})}\BibitemShut {NoStop}%
\bibitem [{\citenamefont {Mathias}\ \emph {et~al.}(2010)\citenamefont
  {Mathias}, \citenamefont {Ruffing}, \citenamefont {Deicke}, \citenamefont
  {Wiesenmayer}, \citenamefont {Sakar}, \citenamefont {Bihlmayer},
  \citenamefont {Chulkov}, \citenamefont {Koroteev}, \citenamefont {Echenique},
  \citenamefont {Bauer} \emph {et~al.}}]{Mathias10}%
  \BibitemOpen
  \bibfield  {author} {\bibinfo {author} {\bibfnamefont {S.}~\bibnamefont
  {Mathias}}, \bibinfo {author} {\bibfnamefont {A.}~\bibnamefont {Ruffing}},
  \bibinfo {author} {\bibfnamefont {F.}~\bibnamefont {Deicke}}, \bibinfo
  {author} {\bibfnamefont {M.}~\bibnamefont {Wiesenmayer}}, \bibinfo {author}
  {\bibfnamefont {I.}~\bibnamefont {Sakar}}, \bibinfo {author} {\bibfnamefont
  {G.}~\bibnamefont {Bihlmayer}}, \bibinfo {author} {\bibfnamefont
  {E.}~\bibnamefont {Chulkov}}, \bibinfo {author} {\bibfnamefont {Y.~M.}\
  \bibnamefont {Koroteev}}, \bibinfo {author} {\bibfnamefont {P.}~\bibnamefont
  {Echenique}}, \bibinfo {author} {\bibfnamefont {M.}~\bibnamefont {Bauer}},
  \emph {et~al.},\ }\href@noop {} {\bibfield  {journal} {\bibinfo  {journal}
  {Phys. Rev. Lett.}\ }\textbf {\bibinfo {volume} {104}},\ \bibinfo {pages}
  {066802} (\bibinfo {year} {2010})}\BibitemShut {NoStop}%
\bibitem [{\citenamefont {Ishizaka}\ \emph {et~al.}(2011)\citenamefont
  {Ishizaka}, \citenamefont {Bahramy}, \citenamefont {Murakawa}, \citenamefont
  {Sakano}, \citenamefont {Shimojima}, \citenamefont {Sonobe}, \citenamefont
  {Koizumi}, \citenamefont {Shin}, \citenamefont {Miyahara}, \citenamefont
  {Kimura} \emph {et~al.}}]{Ishizaka11}%
  \BibitemOpen
  \bibfield  {author} {\bibinfo {author} {\bibfnamefont {K.}~\bibnamefont
  {Ishizaka}}, \bibinfo {author} {\bibfnamefont {M.}~\bibnamefont {Bahramy}},
  \bibinfo {author} {\bibfnamefont {H.}~\bibnamefont {Murakawa}}, \bibinfo
  {author} {\bibfnamefont {M.}~\bibnamefont {Sakano}}, \bibinfo {author}
  {\bibfnamefont {T.}~\bibnamefont {Shimojima}}, \bibinfo {author}
  {\bibfnamefont {T.}~\bibnamefont {Sonobe}}, \bibinfo {author} {\bibfnamefont
  {K.}~\bibnamefont {Koizumi}}, \bibinfo {author} {\bibfnamefont
  {S.}~\bibnamefont {Shin}}, \bibinfo {author} {\bibfnamefont {H.}~\bibnamefont
  {Miyahara}}, \bibinfo {author} {\bibfnamefont {A.}~\bibnamefont {Kimura}},
  \emph {et~al.},\ }\href@noop {} {\bibfield  {journal} {\bibinfo  {journal}
  {Nat. Mater}\ }\textbf {\bibinfo {volume} {10}},\ \bibinfo {pages} {521}
  (\bibinfo {year} {2011})}\BibitemShut {NoStop}%
\bibitem [{\citenamefont {Reeg}\ and\ \citenamefont {Maslov}(2017)}]{Reeg17}%
  \BibitemOpen
  \bibfield  {author} {\bibinfo {author} {\bibfnamefont {C.}~\bibnamefont
  {Reeg}}\ and\ \bibinfo {author} {\bibfnamefont {D.~L.}\ \bibnamefont
  {Maslov}},\ }\href@noop {} {\bibfield  {journal} {\bibinfo  {journal} {Phys.
  Rev. B}\ }\textbf {\bibinfo {volume} {95}},\ \bibinfo {pages} {205439}
  (\bibinfo {year} {2017})}\BibitemShut {NoStop}%
\bibitem [{\citenamefont {Datta}\ and\ \citenamefont {Das}(1990)}]{Datta90}%
  \BibitemOpen
  \bibfield  {author} {\bibinfo {author} {\bibfnamefont {S.}~\bibnamefont
  {Datta}}\ and\ \bibinfo {author} {\bibfnamefont {B.}~\bibnamefont {Das}},\
  }\href@noop {} {\bibfield  {journal} {\bibinfo  {journal} {Appl. Phys.
  Lett.}\ }\textbf {\bibinfo {volume} {56}},\ \bibinfo {pages} {665} (\bibinfo
  {year} {1990})}\BibitemShut {NoStop}%
\bibitem [{\citenamefont {Z{\"u}licke}\ and\ \citenamefont
  {Schroll}(2001)}]{Zulicke01}%
  \BibitemOpen
  \bibfield  {author} {\bibinfo {author} {\bibfnamefont {U.}~\bibnamefont
  {Z{\"u}licke}}\ and\ \bibinfo {author} {\bibfnamefont {C.}~\bibnamefont
  {Schroll}},\ }\href@noop {} {\bibfield  {journal} {\bibinfo  {journal} {Phys.
  Rev. Lett.}\ }\textbf {\bibinfo {volume} {88}},\ \bibinfo {pages} {029701}
  (\bibinfo {year} {2001})}\BibitemShut {NoStop}%
\bibitem [{\citenamefont {Taguchi}\ \emph {et~al.}(2017)\citenamefont
  {Taguchi}, \citenamefont {Zhou}, \citenamefont {Kawaguchi}, \citenamefont
  {Tanaka},\ and\ \citenamefont {Law}}]{Taguchi17}%
  \BibitemOpen
  \bibfield  {author} {\bibinfo {author} {\bibfnamefont {K.}~\bibnamefont
  {Taguchi}}, \bibinfo {author} {\bibfnamefont {B.~T.}\ \bibnamefont {Zhou}},
  \bibinfo {author} {\bibfnamefont {Y.}~\bibnamefont {Kawaguchi}}, \bibinfo
  {author} {\bibfnamefont {Y.}~\bibnamefont {Tanaka}}, \ and\ \bibinfo {author}
  {\bibfnamefont {K.~T.}\ \bibnamefont {Law}},\ }\href@noop {} {\  (\bibinfo
  {year} {2017})}\BibitemShut {NoStop}%
\bibitem [{\citenamefont {Hasan}\ and\ \citenamefont {Kane}(2010)}]{Hasan10}%
  \BibitemOpen
  \bibfield  {author} {\bibinfo {author} {\bibfnamefont {M.~Z.}\ \bibnamefont
  {Hasan}}\ and\ \bibinfo {author} {\bibfnamefont {C.~L.}\ \bibnamefont
  {Kane}},\ }\href {\doibase 10.1103/RevModPhys.82.3045} {\bibfield  {journal}
  {\bibinfo  {journal} {Rev. Mod. Phys.}\ }\textbf {\bibinfo {volume} {82}},\
  \bibinfo {pages} {3045} (\bibinfo {year} {2010})}\BibitemShut {NoStop}%
\bibitem [{\citenamefont {{Hasan}}\ and\ \citenamefont
  {{Moore}}(2011)}]{Hasan11}%
  \BibitemOpen
  \bibfield  {author} {\bibinfo {author} {\bibfnamefont {M.~Z.}\ \bibnamefont
  {{Hasan}}}\ and\ \bibinfo {author} {\bibfnamefont {J.~E.}\ \bibnamefont
  {{Moore}}},\ }\href {\doibase 10.1146/annurev-conmatphys-062910-140432}
  {\bibfield  {journal} {\bibinfo  {journal} {Ann. Rev. Cond. Mat. Phys.}\
  }\textbf {\bibinfo {volume} {2}},\ \bibinfo {pages} {55} (\bibinfo {year}
  {2011})}\BibitemShut {NoStop}%
\bibitem [{\citenamefont {Manchon}\ \emph {et~al.}(2015)\citenamefont
  {Manchon}, \citenamefont {Koo}, \citenamefont {Nitta}, \citenamefont
  {Frolov},\ and\ \citenamefont {Duine}}]{Manchon15}%
  \BibitemOpen
  \bibfield  {author} {\bibinfo {author} {\bibfnamefont {A.}~\bibnamefont
  {Manchon}}, \bibinfo {author} {\bibfnamefont {H.~C.}\ \bibnamefont {Koo}},
  \bibinfo {author} {\bibfnamefont {J.}~\bibnamefont {Nitta}}, \bibinfo
  {author} {\bibfnamefont {S.~M.}\ \bibnamefont {Frolov}}, \ and\ \bibinfo
  {author} {\bibfnamefont {R.~A.}\ \bibnamefont {Duine}},\ }\href {\doibase
  10.1038/NMAT4360} {\bibfield  {journal} {\bibinfo  {journal} {Nature
  material}\ } (\bibinfo {year} {2015}),\ 10.1038/NMAT4360}\BibitemShut
  {NoStop}%
\end{thebibliography}
\end{document}